\documentclass[%
 aip,
 jmp,%
 amsmath,amssymb,
 reprint,%
 nofootinbib,
floatfix,
]{revtex4-1}

\usepackage{amsmath,amssymb,bm,bbm,mathrsfs,amscd}
\usepackage{dsfont}
\usepackage{graphicx}
\usepackage{epstopdf}
\usepackage{epsfig}
\usepackage{tikz}
\usepackage{enumerate,pgfplots}
\usepackage{algpseudocode}
\pgfplotsset{ 
  compat=newest, 
  legend style =
  {font=\footnotesize \sffamily},
  label style = {font=\small\sffamily},
every tick label/.append style={font=\small}}
\usetikzlibrary {positioning}
\usepgfplotslibrary{fillbetween}
\usepackage{amsthm}
\usepackage{nomencl}
\usepackage{braket}
\usepackage{cancel}
\usepackage{mathbbol}
\usepackage{enumitem}
\usepackage{appendix}
\usepackage{placeins}
\usepackage{multirow}
\usepackage[all]{nowidow}

\usepackage[normalem]{ulem}

\renewcommand{\vec}{\bm}

\newcommand{\dd}{\mathrm{d}}
\newcommand{\ii}{\mathrm{i}}

\newcommand{\grad}{\vec{\nabla}}
\newcommand{\p}{\partial}

\newcommand{\cc}{^{\star}}
\newcommand{\inv}{^{-1}}

\newcommand{\libxc}{lib\_xc}

\newcommand{\dt}{\delta t}

\newcommand{\Nd}{N_{\text{D}}}

\newcommand{\Ne}{N_{\text{el}}}

\newcommand{\angstrom}{\mbox{\normalfont\AA}}
\newcommand{\hartree}{\mbox{\normalfont $E_\mathrm{h}$}}

\newcommand{\um}[1]{\ensuremath{\mathrm{\,#1}}} 

\DeclareSymbolFontAlphabet{\mathbbm}{bbold}
\DeclareSymbolFontAlphabet{\mathbb}{AMSb}

\DeclareSymbolFont{newfont}{OML}{cmm}{m}{it}
\DeclareMathSymbol{\Epsilon}{3}{newfont}{15}
\DeclareMathSymbol{\Varrho}{3}{newfont}{37}

\DeclareMathOperator{\FFT}{FT}
\DeclareMathOperator{\iFFT}{IFT}

\begin{document}

\preprint{AIP/123-QED}

\title[]{Mass-Zero constrained dynamics for simulations based on orbital-free density functional theory}

\author{A. Coretti}
 \affiliation{Faculty of Physics, University of Vienna, Kolingasse 14-16, 1090 Vienna, Austria}
 \affiliation{Department of Mathematical Sciences, Politecnico di Torino, I-10129 Torino, Italy}
 \affiliation{Centre Europ\'een de Calcul Atomique et Mol\'eculaire (CECAM), Ecole Polytechnique F\'ed\'erale de Lausanne, 1015 Lausanne, Switzerland}
 \author{T. Baird}
 \affiliation{Centre Europ\'een de Calcul Atomique et Mol\'eculaire (CECAM), Ecole Polytechnique F\'ed\'erale de Lausanne, 1015 Lausanne, Switzerland}
\author{R. Vuilleumier}%
\affiliation{ 
PASTEUR, D\'epartement de chimie, \'Ecole normale sup\'erieure, PSL University, Sorbonne Universit\'e. CNRS, 75005 Paris, France
}%
\author{S. Bonella}
 \email{sara.bonella@epfl.ch}
 \affiliation{Centre Europ\'een de Calcul Atomique et Mol\'eculaire (CECAM), Ecole Polytechnique F\'ed\'erale de Lausanne, 1015 Lausanne, Switzerland}

\date{\today}

\begin{abstract}A new algorithm for efficient and fully time-reversible integration of {\it first-principles} molecular dynamics based on orbital-free density functional theory (OFDFT) is presented. The algorithm adapts to this nontrivial case the recently introduced Mass-Zero (MaZe) constrained dynamics. The formalism ensures that full adiabatic separation is enforced between nuclear and electronic degrees of freedom and, consequently, that the exact Born-Oppenheimer probability for the nuclei is sampled. Numerical integration of the MaZe dynamics combines standard molecular dynamics algorithms, e.g. Verlet or velocity Verlet, with the SHAKE method to impose the minimum conditions on the electronic degrees of freedom as a set of constraints. The developments presented in this work, that include a bespoke adaptation of the standard SHAKE algorithm, ensure that the quasilinear scaling of OFDFT is preserved by the new method 
for a broad range of kinetic and exchange-correlation functionals, including nonlocal ones.
The efficiency and accuracy of the approach is demonstrated via calculations of static and dynamic properties of liquid sodium in the constant energy and constant temperature ensembles.
\end{abstract}

\maketitle

\section{Introduction}
In this paper, we detail the use of the Mass-Zero (MaZe) constrained dynamics for the interesting case of \textit{first-principles} molecular dynamics based on Orbital-Free Density Functional Theory (OFDFT). 

MaZe is a general method for simulating the evolution of a set of physical degrees of freedom (dofs) whose dynamics depends on parameters subject to given conditions.\cite{coretti:2018b,bonella:2020} In MaZe these parameters are treated as degrees of freedom of zero inertia in an extended dynamical system and the conditions are imposed as constraints. The method is formulated in the Lagrangian framework, enabling the key properties of the approach, namely full time reversibility and exact sampling of the probability density of the physical dofs, to emerge naturally from a fully consistent dynamical and statistical viewpoint. An important application domain for MaZe is adiabatic dynamics. In the following we shall focus, in particular, on the \textit{first-principles} adiabatic evolution of a set of coupled ionic (slow) and electronic (fast) dofs. The energy of this system is a function of the ionic positions and a functional of the electronic density. In OFDFT-based simulations, the density is represented in a basis, e.g. plane waves, and the expansion coefficients play the role of the electronic dofs. The well-separated timescales of the motion of the slow and fast dofs enable one to adopt the Born-Oppenheimer approximation for the dynamics. In this approximation, classical evolution equations govern the ionic motion, but the forces acting on these dofs depend parametrically on the values of the fast variables. These values, in turn, are determined by the condition that the energy is at a minimum with respect to the fast dofs for each ionic configuration along the trajectory. 

Typical numerical methods for \textit{first-principles} MD can be grouped into two main families. The first directly implements Born-Oppenheimer dynamics: standard classical MD for the ions is combined with minimization of the energy with respect to the electronic dofs, for example via nonlinear conjugate gradient~\cite{aguado:2003a,jahn:2004} or via \textit{ad hoc} evolutions as in the always stable predictor-corrector method.\cite{kolafa:2004,genzer:2004} The second implements Car-Parrinello MD:\cite{car:1985} the fast dofs are treated as dynamical variables (of finite mass) by defining an appropriate Lagrangian and the minimum is approximately tracked by the overall dynamics.\cite{pastore:1991} A hybrid method that introduces new dynamical variables, in addition to those directly associated with the electronic dofs, and uses them to provide good starting values for energy minimizers such as the conjugate-gradient method, was also proposed in the extended Lagrangian Born-Oppenheimer (XL-BOMD) framework.\cite{niklasson:2006,niklasson:2020,niklasson:2021} These methods are routinely used and have produced remarkable results, but they suffer from problems that motivate further developments. In particular, full convergence of minimization based approaches can be too expensive for large-scale simulations of condensed-phase systems.\cite{pounds:2009} Incomplete convergence has serious consequences on the evolution, including instabilities,\cite{pacaud:2018} violation of adiabaticity~\cite{remler:1990,pulay:2004} and therefore energy drifts and breaking of time reversibility of the propagation. Violation of adiabaticity affects also Car-Parrinello MD~\cite{aguado:2003a} and forces either the use of thermostats (which, however, lead to questions on the statistical ensemble sampled) or of a very small mass associated to the auxiliary variables (which, however, imposes a very small timestep for the numerical propagation). Full time reversibility or ease of portability of the algorithm is problematic for the predictor-corrector and XL-BOMD methods, that have system dependent parameters or forces that must be tuned for each calculation.\cite{kolafa:2004,genzer:2004} Indeed, the violation of strict adiabaticity is intrinsic to extended-Lagrangian methods as long as the mass of the auxiliary variables is finite.\cite{payne:1986,pastore:1991,payne:1992,marx:2012-book}

MaZe provides an alternative approach that eliminates or mitigates the problems mentioned above. It employs the framework of constrained dynamics~\cite{goldstein:2002-book,ciccotti:1986} to ensure fully adiabatic propagation and, consequently, correct sampling of the target probability for the physical dofs. Using an adapted SHAKE algorithm,\cite{ryckaert:1977} it enables efficient satisfaction of the minimum condition with a level of precision close to the numerical limit.\cite{coretti:2018b,bonella:2020,coretti:2020a} The SHAKE-based integration algorithm is symplectic~\cite{leimkuhler:2015-book} and fully time reversible, leading to long-time stability of the simulation with a timestep determined by the typical frequencies of the motion of the slow dofs. Recent applications of MaZe that illustrate these properties cover the simulation of classical models for polarizable systems,\cite{coretti:2018b,coretti:2020a,marin-lafleche:2020} including the nontrivial case of classical polarizable systems in external magnetic field,\cite{girardier:2021} and proof-of-principle OFDFT calculations that employed simple energy functionals.\cite{bonella:2020}

In the following, we considerably extend our previous exploratory work for MaZe in OFDFT simulations by tackling two important issues. Firstly, as discussed in detail in Sec.~\ref{sec:MaZe_OFDFT}, the MaZe OFDFT constrained variables are the \textit{complex} coefficients in a Fourier-series expansion of the electronic density. This implies that the Lagrange multipliers associated to our problem are also complex. The generalization of the SHAKE algorithm to this case is presented in this work. The second, critical problem that we address stems from the specific nature of the minimum constraint. As a rule, SHAKE has been applied to relatively simple constraints, involving only small subsets of degrees of freedom, e.g. fixing bond lengths in molecules where the functional form of the constraint is quadratic in the coordinates of the few atoms involved in each bond. This local structure is key for the rapid convergence of the iterative algorithm (almost) universally adopted for solving the constrained evolution of the system.\cite{ryckaert:1977} The minimum constraint appearing in MaZe, on the other hand, is typically a global and highly nonlinear function of many auxiliary degrees of freedom that, for general energy functionals, cannot be solved efficiently with the standard algorithm. In this work then, we introduce an alternative procedure, based on work of Weinbach and Elber,\cite{weinbach:2005} to solve the constraint equations and prove its remarkable efficacy. This development is essential in view of MaZe deployment to large-scale simulations based on OFDFT with functionals of arbitrary complexity, including nonlocal kinetic and exchange-correlation functionals.\cite{wang:1992,wang:1999}

The paper is organized as follows. We start by discussing in detail the MaZe formalism for the specific case of OFDFT \textit{first-principles} MD at constant number of particles, volume and energy. We then show how the constrained dynamics is adapted to the case of complex Lagrange multipliers. In Sec.~\ref{sec:Algorithm}, we discuss first the solution of the constraints via the standard algorithm (see Sec.~\ref{subsec:StandardSHAKE}) and highlight its high numerical cost for generic kinetic and exchange-correlation functionals. In Sec.~\ref{subsec:ElberSHAKE} the new, numerically efficient and general algorithm for the solution of the constraint equations is then introduced. 
The accuracy and efficiency enabled by these developments are illustrated in the Results section for constant energy and constant temperature systems. Here, we first validate the algorithm via benchmark calculations on liquid sodium and then compare the numerical performance (both in terms of number of operations per timestep and in clocking of the timestep) of the new and standard SHAKE algorithms. In particular, we demonstrate that, for commonly adopted energy functionals at the GGA level, the algorithm scales quasilinearly with system size. 

\section{MaZe dynamics for orbital-free DFT}\label{sec:MaZe_OFDFT}
OFDFT was proposed\cite{pearson:1993} as a computationally convenient alternative to DFT dynamics based on the use of Kohn-Sham orbitals.~\cite{marx:2012-book} The main advantage of OFDFT is that, in contrast to Kohn-Sham-based DFT, it enables a representation of the total energy that does not require auxiliary orthonormal orbitals. This entails that the overall cost of the calculation per timestep is quasilinear in the system's volume $\Omega$, ($\mathcal{O}(\Omega\log\Omega)$) and memory requirements conveniently scale linearly as $\mathcal{O}(\Omega)$. Unfortunately, developing reliable and transferable kinetic-energy functionals for general systems (and for different phases of matter) is an open challenge~\cite{ligneres:2005-inbook} that prevents universal use of the method. In spite of this well-known issue, OFDFT remains a useful approach, often enabling the simulation of system sizes out of reach of alternative methods.\cite{ligneres:2005-inbook,witt:2018} The method is also implemented in different high performance codes for the simulation of materials, such as GPAW~\cite{mortensen:2005,Enkovaara:2010} or PROFESS.\cite{chen:2015} Very recently, the use of machine learning methods to obtain flexible and accurate kinetic-energy functionals~\cite{meyer:2020,ghasemi:2021} has also led to a renewed interest in OFDFT. 

In OFDFT,  the ground state electronic energy is written as a function of the $N$ ionic positions, $\vec{R}\equiv\{\vec{R}_1,...,\vec{R}_N\}$, and as a functional of the electronic density $n$
\begin{equation}
\label{eq:EnergyODFT}
\begin{array}{cc}
 E[\vec{R}, n]&  \equiv E_{\text{KE}}[n]+E_{\text{Hart}}[n]+E_{\text{xc}}[n] \\
  &+ E_{\text{ext}}[\vec{R},n] + E_{\text{Coul}}(\vec{R}),
\end{array}
\end{equation}
where $E_{\text{KE}}[n]$ is the kinetic energy of a noninteracting electron gas at the density of the physical system
, and $E_{\text{Hart}}[n]$ and $E_{\text{xc}}[n]$ are the Hartree and exchange-correlation energies, respectively. $E_{\text{Coul}}(\vec{R})$ is the classical Coulomb energy of the ions and $E_{\text{ext}}[\vec{R},n]$ is the interaction energy between the electrons and the ions. In MD simulations, the density is typically represented in the plane-wave basis 
\begin{equation}
n(\vec{r})=\sum_{\vec{G}}\tilde{n}_{\vec{G}}e^{i\vec{G}\cdot\vec{r}},
\end{equation}
where $\vec{G}=\frac{2\pi}{L}(i,j,k)$, with $L$ the size of the cubic box and $i$, $j$, $k$ integers. This expansion is truncated with $\tilde{n}_{\vec{G}}=0$ if any component of the wavevector is larger than a cutoff, $\vert G^l \vert > G_\text{c}$, $l=x,y\text{ or }z$.  Since the density in real space $n(\vec{r})$ is,  by definition,  a real function,  the Fourier coefficients $\tilde{n}_{\vec{G}}$ (typically complex quantities),  satisfy the relationship $\tilde{n}_{-\vec{G}}=\tilde{n}\cc_{\vec{G}}$.  Once the Fourier representation is adopted,  the ground state energy becomes a function of the ionic positions and the expansion coefficients.  

\textit{First-principles} OFDFT molecular dynamics relies on the Born-Oppenheimer approximation: the forces on the ions are computed at each timestep as the gradient of the ground state energy with respect to the ionic positions, with the energy computed at an electronic density that minimizes its value.  In the plane wave representation of the density, this minimum condition is equivalent to imposing that the derivatives of the ground state energy with respect to the coefficients of the Fourier expansion are null.  In taking these derivatives,  we have the choice to adopt as independent variables the real and imaginary parts of $\tilde{n}_{\vec{G}}$ or to consider instead $\tilde{n}_{\vec{G}}$ and $\tilde{n}\cc_{\vec{G}}$.  Indeed,  the definition of the complex derivatives (also known as Wirtinger derivatives)
\begin{equation}
\begin{aligned}
    \frac{\partial E}{\partial \tilde{n}\cc_{\vec{G}}}&=\frac{1}{2}\left(
    \frac{\partial E}{\partial \text{Re}[\tilde{n}_{\vec{G}}]}
    +\ii\frac{\partial E}{\partial \text{Im}[\tilde{n}_{\vec{G}}]}
    \right),\\
    \frac{\partial E}{\partial \tilde{n}_{\vec{G}}}&=\frac{1}{2}\left(
    \frac{\partial E}{\partial \text{Re}[\tilde{n}_{\vec{G}}]}
    -\ii\frac{\partial E}{\partial \text{Im}[\tilde{n}_{\vec{G}}]}
    \right),
\end{aligned}
\end{equation}
(where $\text{Re}[\tilde{n}_{\vec{G}}]$ and $\text{Im}[\tilde{n}_{\vec{G}}]$ are the real and imaginary parts of $\tilde{n}_{\vec{G}}$) ensures the equivalence of these choices. In the following,  we adopt  $\tilde{n}_{\vec{G}}$ and $\tilde{n}\cc_{\vec{G}}$ as independent variables. Furthermore,  from now on all Fourier coefficients are taken in the interval $\vec{G}\in [\vec{0},\vec{G}_{\text{max}}]$, a shorthand notation to indicate that wavevectors span one-half space (we denote $N_G$ the number of such vectors). With these choices, the Born-Oppenheimer conditions become
\begin{equation}
\label{eq:BO_cond_recip}
\begin{aligned}
\sigma_{\vec{G}_\alpha}(\vec{R},\tilde{\vec{n}},  \tilde{n}_{\vec{G}_0}) = \frac{\p E(\vec{R}, \tilde{\vec{n}},\tilde{n}_{\vec{G}_0})}{\p \tilde{n}\cc_{\vec{G}_\alpha}} = 0 \ \ \text{for} \ \ \alpha \in [1,N_G]; \\
\sigma\cc_{\vec{G}_\alpha}(\vec{R},\tilde{\vec{n}},\tilde{n}_{\vec{G}_0}) = \frac{\p E(\vec{R}, \tilde{\vec{n}},\tilde{n}_{\vec{G}_0})}{\p \tilde{n}_{\vec{G}_\alpha}} = 0 \ \ \text{for} \ \ \alpha \in [0,N_G].
\end{aligned}
\end{equation}
In the equation above,  we have introduced, for future convenience,  the vector 
\begin{equation}
\tilde{\vec{n}}=\{\tilde{n}_{\vec{G}_1},\ldots,\tilde{n}_{\vec{G}_{N_G}},\tilde{n}\cc_{\vec{G}_{1}},\ldots,\tilde{n}\cc_{\vec{G}_{N_G}}\},
\end{equation}
that contains the $2N_G$ Fourier coefficients other than the one corresponding to $\vec{G}_0=(0,0,0)$. This latter coefficient is associated to the additional condition that the total number of electrons $\Ne$, must be conserved along the dynamics. In the plane wave basis, this condition is expressed as
\begin{equation}\label{eq:TotEl}
\tilde{n}_{\vec{G}_0}\Omega = \Ne,
\end{equation}
where $\Omega$ is the volume of the simulation box.

The MaZe evolution for OFDFT \textit{first-principles} dynamics is obtained by promoting the coefficients $\tilde{n}_{\vec{G}_\alpha}$ and $\tilde{n}\cc_{\vec{G}_\alpha}$ to the role of dynamical variables. The evolution equations, obtained from a Lagrangian formulation of the dynamics of the ions and these additional variables, are then complemented with the conditions in Eqs.~\eqref{eq:BO_cond_recip} and~\eqref{eq:TotEl}, which are interpreted as constraints. In the following, we impose the conservation of the number of electrons by enforcing Eq.~\eqref{eq:TotEl} and the minimum condition on $\tilde{n}_{\vec{G}_0}$ as an initial condition. We then exclude this quantity from the set of auxiliary dynamical variables and consider it as a constant in the simulations. To simplify the notation, we also drop the explicit indication of the dependence of the ground state energy and of the constraints on $\tilde{n}_{\vec{G}_0}$.

To proceed, we assign (temporarily) a finite (scalar) inertia, $\mu$, to the additional variables $\tilde{\vec{n}}$, and define the following Lagrangian for the system\footnote{We define the scalar product of two complex vectors, $\vec{a}, \vec{b} \in \mathbb{C}^N$ as $\vec{a}\cc \cdot \vec{b} = \sum \limits_{i=1}^{N} a_i\cc b_i$.}
\begin{equation}
\label{eq:OFDFT_Lagr}
L = \frac{1}{2}\sum_{I}^{N}M_I\dot{\vec{R}}^2_I + \frac{\mu}{2}\dot{\vec{\tilde{n}}}\cc \cdot \dot{\vec{\tilde{n}}}
- E(\vec{R}, \tilde{\vec{n}}).
\end{equation}
We then introduce the vector of the $2N_G$ constraints
\begin{equation}\label{eq:ConstraintVec}
     \vec{\sigma}=\{\sigma_{\vec{G}_1},\ldots,\sigma_{\vec{G}_{N_G}},\sigma\cc_{\vec{G}_{1}},\ldots,\sigma\cc_{\vec{G}_{N_G}}\},
\end{equation}
and the $2N_G$ Lagrange multipliers
\begin{equation}\label{eq:GammaVec}
\vec{\lambda}=\{\lambda_{\vec{G}_1},\ldots\lambda_{\vec{G}_{N_G}},\lambda\cc_{\vec{G}_{1}},\ldots,\lambda\cc_{\vec{G}_{N_G}} \},
\end{equation}
in order to form the constrained Lagrangian
\begin{equation}
\label{eq:OFDFT_Lagr_const}
\mathcal{L} = L + \vec{\sigma}\cc  \cdot \vec{\lambda} .
\end{equation}

Note that, at variance with typical situations in molecular dynamics,  the constraints and the Lagrange multipliers are, in general, \textit{complex} quantities, making it necessary to generalize standard algorithms for their treatment. The prescribed form for the Lagrangian respects two important properties. The constraint term, in fact, can be rewritten as
\begin{equation}
\label{eq:lam_sc_sig}
    \vec{\sigma}\cc \cdot \vec{\lambda} =\text{Re}\left(\frac{\partial E}{\partial \tilde{\vec{n}}}\right) \cdot \text{Re}\left( \vec{\lambda}\right)  + \text{Im}\left(\frac{\partial E}{\partial \tilde{\vec{n}}}\right) \cdot \text{Im}( \vec{\lambda}) ,
\end{equation} where we have introduced the vector of derivatives: 
\begin{align}
    &\frac{\partial E}{\partial \tilde{\vec{n}}}=\left\{\frac{\partial E}{\partial \tilde{n}_{\vec{G}_{1}}},\ldots,\frac{\partial E}{\partial \tilde{n}_{\vec{G}_{N_G}}},\frac{\partial E}{\partial \tilde{n}\cc_{\vec{G}_{1}}},\ldots,\frac{\partial E}{\partial \tilde{n}\cc_{\vec{G}_{N_G}}}\right\}.
\end{align}
Eq.~\eqref{eq:lam_sc_sig} shows that the scalar product of the constraints and Lagrange multipliers is real and, therefore,  so is the constrained Lagrangian,  Eq.~\eqref{eq:OFDFT_Lagr}.  Furthermore,  the expression verifies, once more, the equivalence of choosing $\tilde{n}_{\vec{G}_{\alpha}}$ and $\tilde{n}\cc_{\vec{G}_{\alpha}}$ or the real and imaginary parts of the density coefficients as independent variables.  Indeed,  with the latter choice,  the equation indicates that,  as expected and consistently,  the constraints
\begin{equation}
    \frac{\partial E}{\partial \text{Re}[\tilde{n}_{\vec{G}_{\alpha}}]}=0;\qquad
    \frac{\partial E}{\partial \text{Im}[\tilde{n}_{\vec{G}_{\alpha}}]}=0,
\end{equation}
would be satisfied, via the --- real --- Lagrange multipliers $\text{Re}\left( \lambda_{\vec{G}_{\alpha}}\right)$ and $\text{Im}\left({\lambda}_{\vec{G}_{\alpha}}\right)$.

The constrained evolution equations for the MaZe OFDFT system are then given by
\begin{equation}
\begin{aligned}
M_I\ddot{\vec{R}}_I &= -\nabla_{\vec{R}_I}E(\vec{R}, \tilde{\vec{n}}) 
+    ( \nabla_{\vec{R}_I} \vec{\sigma}\cc ) \cdot \vec{\lambda};  \\
\mu\ddot{{\tilde{\vec{n}}} }&= - \frac{\p E(\vec{R},\tilde{\vec{n}})}{\p \tilde{\vec{n}}\cc} 
+  \frac{\p \vec{\sigma}\cc}{\p \tilde{\vec{n}}\cc} \cdot \vec{\lambda} \\
& \equiv  \frac{\p E(\vec{R},\tilde{\vec{n}})}{\p \tilde{\vec{n}}\cc} + \Sigma^\dagger\cdot \vec{\lambda}, 
\end{aligned}
\end{equation}
where we have introduced the matrix $\Sigma^\dagger\equiv \frac{\p \vec{\sigma}\cc}{\p \tilde{\vec{n}}\cc}$, which is the Hermitian conjugate of 
\begin{equation}\label{eq:SigmaMatrix}
\begin{aligned}
\{\Sigma\}_{\alpha,\beta}=\{\Sigma\}_{N_G+\alpha,N_G+\beta}\cc&=\frac{\p \sigma_{\vec{G}_{\alpha}}}{\p \tilde{n}_{\vec{G}_{\beta}}}; \\
\{\Sigma\}_{\alpha,N_G+\beta}=\{\Sigma\}_{N_G+\alpha,\beta}\cc&=\frac{\p \sigma_{\vec{G}_{\alpha}}}{\p \tilde{n}\cc_{\vec{G}_{\beta}}}.
\end{aligned}
\end{equation}
The evolution equations for the additional variables can be rearranged by first observing that the vector derivative of the energy with respect to the expansion coefficients is the null vector since it corresponds to the constraints. Dividing the evolution equation for the additional variables by the inertia we obtain
\begin{equation}
\begin{aligned}
M_I\ddot{\vec{R}}_I &= -\nabla_{\vec{R}_I}E(\vec{R}, \tilde{\vec{n}}) 
+   ( \nabla_{\vec{R}_I} \vec{\sigma}\cc) \cdot \vec{\lambda};   \\
\ddot{\tilde{\vec{n}}} &= \frac{1}{\mu}
  \Sigma^\dagger \cdot \vec{\lambda}.
\end{aligned}
\end{equation}
The key step in the MaZe approach follows by considering the limit $\mu \to 0$ for the second equation above and imposing that this limit is taken keeping the accelerations of $\tilde{\vec{n}}$ finite. This implies that, for every component of the vector of the Lagrange multipliers, the ratio $\gamma_{\vec{G}_{\beta}} = \lim_{\mu \to0}\frac{\lambda_{\vec{G}_{\beta}}}{\mu}$ must remain finite, entailing that the Lagrange multipliers are proportional to $\mu$. In the limit of zero mass for the auxiliary variables, then, the constraint forces acting on the ionic degrees of freedom vanish and the system becomes
\begin{equation}
\label{eq:OFDFT_MaZe}
\begin{aligned}
M_I\ddot{\vec{R}}_I &= -\nabla_{\vec{R}_I}E(\vec{R}, \tilde{\vec{n}}); \\
\ddot{\tilde{\vec{n}}}&= \Sigma^\dagger\cdot \vec{\gamma}.
\end{aligned}
\end{equation}
with $\vec{\gamma} = \lim_{\mu \to0}\frac{\vec{\lambda}}{\mu}$.

Eq.~\eqref{eq:OFDFT_MaZe} defines the MaZe dynamics for OFDFT. Some important properties must be noted. Firstly, since the Lagrange multipliers $\vec{\lambda}$ go to zero with the auxiliary mass $\mu$, the evolution of the physical variables is not affected by the constraint forces. Secondly, the evolution of $\tilde{ \vec{n}}$, controlled only by the constraint forces, satisfies by construction the minimum conditions at all times. These conditions are thus automatically fulfilled also in the first of Eq.~\eqref{eq:OFDFT_MaZe} which is then fully equivalent to Born-Oppenheimer dynamics. This implies that, by rigorously enforcing the mass-zero limit for the auxiliary variables, MaZe provides an evolution that leads to the exact adiabatic dynamics for the physical variables. This has the important consequence that the probability density sampled from the dynamical system for the ions is exactly the one typically associated with Born-Oppenheimer dynamics.\cite{bonella:2020,coretti:2020a} MaZe is then free of sampling errors typically encountered in schemes with finite values of the auxiliary variables' mass. From the point of view of the algorithm, the first equation can be integrated via any standard MD algorithm (e.g. Verlet) with a timestep determined only by the force acting on the ions. At each timestep, the Lagrange multipliers  $\vec{\gamma}$, that appear as unknown, time-dependent parameters, must also be determined. This is done adopting the SHAKE strategy.~\cite{ryckaert:1977,ciccotti:1986} In the next section, details of the algorithm for OFDFT MaZe dynamics are provided.

\section{MaZe algorithm for orbital-free DFT}\label{sec:Algorithm}

The numerical integration of the dynamical system in Eq.~\eqref{eq:OFDFT_MaZe} proceeds as follows. The evolution of the ionic degrees of freedom is performed via standard classical MD algorithms. Here we discuss the case of Verlet propagation but there is no difficulty in adopting, for example, the velocity version of the scheme. Thus
\begin{equation}
       \vec{R}_I(t+\dt) = 2\vec{R}_I(t)-\vec{R}_I(t-\dt)-\frac{\dt^2}{M_I}\nabla_{\vec{R}_I}E\bigl(\vec{R}(t), \tilde{\vec{n}}(t)\bigr),
\end{equation}
where $\dt$ is the timestep of the simulation. This algorithm requires, at each step, the computation of the forces acting on the ions after advancement of their positions. These forces
\begin{equation}
    \vec{F}_I(t+\dt)=-\nabla_{\vec{R}_I}E\bigl(\vec{R}(t+\dt), \tilde{\vec{n}}(t+\dt)\bigr),
\end{equation}
depend both on the advanced ionic coordinates and on the values of the $\tilde{\vec{n}}(t+\dt)$. The latter can be formally obtained as
\begin{equation}\label{eq:VerletCoeff}
\begin{aligned}
   \tilde{\vec{n}}(t+\dt; \vec{\gamma}) =& 2\tilde{\vec{n}}(t)-\tilde{\vec{n}}(t-\dt) 
   +\dt^2\Sigma(t)^\dagger\cdot\vec{\gamma}  \\
  =& \tilde{\vec{n}}^P + \Sigma(t)^\dagger\cdot\tilde{\vec{\gamma}}.
\end{aligned}
\end{equation}
In the second line above, we have implicitly defined $\tilde{\vec{n}}^P=2\tilde{\vec{n}}(t)-\tilde{\vec{n}}(t-\dt) $, to be called later the ``provisional'' value of the coefficient, and we have scaled the Lagrange multipliers --- as usually done in SHAKE --- by the square of the timestep, so $\tilde{\vec{\gamma}}=\dt^2\vec{\gamma}$.
The updated density coefficients depend on the set of Lagrange multipliers that are, at this stage, unknown. Following the SHAKE strategy, the multipliers are determined by imposing that the constraints 
\begin{equation}\label{eq:FullConstraints}
\begin{aligned}
    \vec{\sigma}\bigl(\vec{R}(t+\dt),\tilde{\vec{n}}(t+\dt;\tilde{\vec{\gamma}})\bigr)&=\vec{0} ,
\end{aligned}
\end{equation}
are satisfied at the positions predicted by the MD algorithm. Eqs.~\eqref{eq:FullConstraints} are, in general, a system of nonlinear equations to be solved for the Lagrange multipliers. The solution of the nonlinear system is obtained via an adapted iterative Newton-Raphson scheme in which the values of the Lagrange multipliers at iteration ${\kappa}+1$ are given by
\begin{equation}\label{eq:NewtonRaph}
    {\tilde{\vec{\gamma}}^{{\kappa}+1}} = {\tilde{\vec{\gamma}}^{\kappa}} - \omega\mathbb{J}\inv(\tilde{\vec{n}}^{\kappa})\vec{\sigma}\bigl(\vec{R}(t+\dt),\tilde{\vec{n}}^{\kappa}\bigr).
\end{equation}
In the equation above, we have introduced $\mathbb{J}\inv(\tilde{\vec{n}}^{\kappa})$,  the inverse of the Jacobian $ \mathbb{J}(\tilde{\vec{n}}^{\kappa})\equiv \frac{\p \vec{\sigma}(\tilde{\vec{n}}^{\kappa})}{\p \tilde{\vec{\gamma}}}$ whose matrix elements are given by
\begin{equation}\label{eq:Jacobian}
\begin{aligned}
\{\mathbb{J}(\tilde{\vec{n}}^{\kappa})\}_{\alpha,\beta}=\{\mathbb{J}(\tilde{\vec{n}}^{\kappa})\}_{N_G+\alpha,N_G+\beta}\cc&=\frac{\p \sigma_{\vec{G}_{\alpha}}(\tilde{\vec{n}}^{\kappa})}{\p \tilde{\gamma}_{\vec{G}_{\beta}}};\\
\{\mathbb{J}(\tilde{\vec{n}}^{\kappa})\}_{\alpha,N_G+\beta}=\{\mathbb{J}(\tilde{\vec{n}}^{\kappa})\}_{N_G+\alpha,\beta}\cc&=\frac{\p \sigma_{\vec{G}_{\alpha}}(\tilde{\vec{n}}^{\kappa})}{\p \tilde{\gamma}\cc_{\vec{G}_{\beta}}},
\end{aligned}
\end{equation}
with $\alpha,\beta=1,...,N_G$, and  $\tilde{\vec{n}}^{\kappa}$ is the vector of the electronic density coefficients at iteration $\kappa$. The parameter $\omega$ in Eq.~\eqref{eq:NewtonRaph} is a scaling factor that, following common practice in minimization algorithms,~\cite{nocedal:2006-book} modulates the update of the Lagrange multipliers. Previous calculations~\cite{barth:1995} show that using such a scaling improves convergence in standard implementations of SHAKE. At each iteration in the cycle, the Fourier coefficients are updated as
\begin{equation}\label{eq:NewRapCoeffUpdate}
\begin{aligned}
{\tilde{\vec{n}}^{\kappa+1}}&={\tilde{\vec{n}}}^{\kappa} + \Sigma(t)^\dagger\cdot\tilde{\vec{\gamma}}^\kappa,
\end{aligned}
\end{equation}
and used in Eq.~\eqref{eq:NewtonRaph} to update the vector of constraints. 
The iteration cycle is typically initialized using null values for the Lagrange multipliers or assigning them as the converged values of the previous timestep. The process is stopped when  $\max_{\vec{G}_{\alpha}} |{\sigma_{\vec{G}_{\alpha}}}|$ becomes smaller than a predefined tolerance. 

The convergence of the Newton-Raphson algorithm is normally quadratic.\cite{nocedal:2006-book} The numerical efficiency of the method therefore hinges on the cost of the different quantities and operations necessary to perform the iterations.  Below, we summarize the main characteristics of these quantities by showing --- when necessary --- their explicit component-wise expression, reporting the numerical cost of the key operations in the algorithm.  Detailed assessment of these numerical costs is given in the Appendices to avoid interrupting the main text with lengthy calculations.

Let us begin with the evaluation of the constraint vector $\vec{\sigma}\bigl(\vec{R}(t+\dt),\tilde{\vec{n}}^\kappa\bigr)$. From Eq.~\eqref{eq:FullConstraints}, we have to compute, for example,
\begin{equation}
\label{eq:DEofnk_dnstar}
\sigma_{\vec{G}_{\alpha}}\bigl(\vec{R}(t+\dt),\tilde{\vec{n}}^{\kappa}\bigr)=\frac{\p E\bigl(\vec{R}(t+\dt), \tilde{\vec{n}}^{\kappa}\bigr)}{\p \tilde{n}\cc_{\vec{G}_{\alpha}}} ,
\end{equation}
where the ground state energy, $E$ is expressed as the sum in Eq.~\eqref{eq:EnergyODFT}. The $\sigma\cc_{\vec{G}_{\alpha}}\bigl(\vec{R}(t+\dt),\tilde{\vec{n}}^{\kappa}\bigr)$ can be obtained via complex conjugation. Eq.~\eqref{eq:DEofnk_dnstar} highlights that the calculation of the constraints is identical to that of the force on the electronic coefficients in Car-Parrinello OFDFT~\cite{pearson:1993} or of the derivatives employed in conjugate-gradient minimization in Born-Oppenheimer OFDFT~\cite{chen:2015}. Direct evaluation shows that the derivatives are obtained at an $\Omega\log\Omega$ cost for any form of the energy functional, and in particular of the kinetic energy functional. We do not report the explicit expressions for these quantities since they are standard and available in the literature.\cite{wang:1999,chen:2015}

Secondly, the evolution equations~\eqref{eq:VerletCoeff} and the update of the density coefficients~\eqref{eq:NewRapCoeffUpdate} require a matrix-vector product  $\Sigma^\dagger \cdot \vec{\gamma}$. 
In Appendix~\ref{app:MV_GGA} we show that this can be performed with $\Omega \log \Omega$ scaling for GGA functionals and in Appendix~\ref{app:MV_nonlocal} that the same scaling can be achieved for nonlocal functionals.  
Having a procedure to compute the matrix-vector product between $\Sigma^\dagger$ and a generic vector, also offers a means to compute the matrix-vector product of a generic vector with $\mathbb{J}$, exploiting the fact that the Jacobian can be expressed as $\mathbb{J}=\Sigma(\tilde{\vec{n}}^{\kappa})\cdot \Sigma(t)^\dagger$. The latter relationship can be proven by considering the definition of the matrix elements in Eq.~\eqref{eq:Jacobian}.  For example,
    \begin{align}
\label{eq:dsigma_dgamma}
        \nonumber
       \{\mathbb{J}&(\tilde{\vec{n}}^{\kappa})\}_{\alpha,\beta}  = \frac{\p \sigma_{\vec{G}_{\alpha}}}{\p \tilde{\gamma}_{\vec{G}_{\beta}}} \\ \nonumber 
      &=\sum^{N_G}_{\lambda=1}\biggl[\frac{\p \sigma_{\vec{G}_{\alpha}}}{\p \tilde{n}_{\vec{G}_{\lambda}}}\frac{\p \tilde{n}_{\vec{G}_{\lambda}}}{\p \tilde{\gamma}_{\vec{G}_{\beta}}} + \frac{\p \sigma_{\vec{G}_{\alpha}}}{\p \tilde{n}\cc_{\vec{G}_{\lambda}}}\frac{\p \tilde{n}\cc_{\vec{G}_{\lambda}}}{\p \tilde{\gamma}_{\vec{G}_{\beta}}} \biggr ] \\\nonumber
        &= \sum^{N_G}_{\lambda=1}\biggl[\frac{\p \sigma_{\vec{G}_{\alpha}}(\tilde{\vec{n}}^{\kappa})}{\p \tilde{n}_{\vec{G}_{\lambda}}}\frac{\p \sigma\cc_{\vec{G}_{\beta}}(t)}{\p \tilde{n}\cc_{\vec{G}_{\lambda}}} + \frac{\p \sigma_{\vec{G}_{\alpha}}(\tilde{\vec{n}}^{\kappa})}{\p \tilde{n}\cc_{\vec{G}_{\lambda}}}\frac{\p \sigma\cc_{\vec{G}_{\beta}}(t)}{\p \tilde{n}_{\vec{G}_{\lambda}}} \biggr ] \\
        &= \left\{ \Sigma(\tilde{\vec{n}}^{\kappa})\cdot \Sigma(t)^\dagger \right\}_{\alpha,\beta}.
    \end{align}
In going from the first to the second line in the equations above,  the derivative of the density coefficients with respect to the Lagrange multipliers was computed from Eq.~\eqref{eq:VerletCoeff} and the notation indicates that derivatives of the constraints are computed either at the provisional values of the coefficients, $\tilde{\vec{n}}^{\kappa}$, or with variables computed at time $t$.  Similar expressions hold for the other blocks of the matrix proving the result. 
Finally, the solution of the Newton-Raphson system, Eq.~\eqref{eq:NewtonRaph}, requires evaluation and inversion of the Jacobian, Eq.~\eqref{eq:Jacobian}. These steps are characteristic of the MaZe approach and nontrivial. The number of elements in the Jacobian matrix, in fact, scales quadratically with the number of electronic coefficients and brute force inversion of the matrix would take the cost of MaZe dynamics to cubic order in CPU operations and to quadratic order in memory requirements. Indeed, it is well known even in standard applications of classical constrained MD, that the dimensions of the Jacobian are typically too large to allow for direct inversion. The commonly adopted strategy is then to consider the diagonal approximation of the Jacobian (i.e., all off-diagonal elements are set to zero) enabling inversion at linear cost in the matrix dimensions. In the next subsection, we discuss in detail this consolidated approach, that we shall indicate as standard SHAKE and that has been employed also in previous work with MaZe. For energy functionals beyond the GGA class, such as those including the more recent and accurate nonlocal kinetic energies, however, linear scaling of this algorithm fails due to the intrinsic cost of computing the diagonal elements of the Jacobian. In the following subsection, we then describe an effective and completely general alternative implementation of SHAKE that preserves the $\Omega\log\Omega$ scaling of OFDFT simulations also for nonlocal functionals. Both implementations of SHAKE discussed in the following avoid propagation of the (numerical) error in the enforcement of the constraint at different timesteps. Furthermore, they are symplectic and time reversible,\cite{leimkuhler:1994,leimkuhler:2004-book} leading to stable propagation for very long times.

Before moving to the details of the SHAKE implementations, we note that the algorithm described so far for a system of ions evolving at constant energy (NVE ensemble) can be applied with very limited changes also to ionic evolution in other ensembles. In particular, ionic motion at constant temperature will be considered in the Results section. In the full adiabatic limit, the overall system is characterized in this case by a steady nonequilibrium regime in which the electronic degrees of freedom remain strictly at zero temperature and respect the minimum conditions, while the ions move at an assigned temperature, $T$. The coupled dynamical system is then obtained by considering thermostatted evolution equations (e.g. Langevin) for the ionic degrees of freedom and constraining the electronic coefficients at each configuration along the ionic trajectory as described above.

\subsection{Standard SHAKE (STD-SHAKE)}\label{subsec:StandardSHAKE}
The diagonal approximation of the Jacobian matrix defined in Eq.~\eqref{eq:Jacobian} is given by
\begin{equation}
\label{eq:diagonal}
\begin{aligned}
    \{\mathbb{J}_D\}_{\alpha,\beta}&=\frac{\p \sigma_{\vec{G}_\alpha}}{\p \tilde{\gamma}_{\vec{G}_\beta}}\delta_{\alpha,\beta};\\
    \{\mathbb{J}_D\}_{\alpha,N_G+\beta}&=0;\\
    \{\mathbb{J}_D\}_{N_G+\alpha,\beta}&=0;\\
    \{\mathbb{J}_D\}_{N_G+\alpha,N_G+\beta}&=\frac{\p \sigma\cc_{\vec{G}_\alpha}}{\p \tilde{\gamma}\cc_{\vec{G}_\beta}}\delta_{\alpha,\beta},
\end{aligned}
\end{equation}
leading to approximate Eq.~\eqref{eq:NewtonRaph} as
\begin{equation}\label{eq:NewtonRaphDiag}
    {\tilde{\vec{\gamma}}^{{\kappa}+1}} = {\tilde{\vec{\gamma}}^{\kappa}} - \omega\mathbb{J}_D\inv(\tilde{\vec{n}}^{\kappa})\vec{\sigma}\bigl(\vec{R}(t+\dt),\tilde{\vec{n}}^{\kappa}\bigr).
\end{equation}
This diagonal approximation trivially decouples the first $N_G$ components of the vectors $\vec{\sigma}$ and $\vec{\gamma}$ from  the second $N_G$ components. Furthermore, as the second $N_G$ equations in the diagonal system corresponds to the complex conjugate of the first $N_G$, we only need to solve for the $N_G$-dimensional vector $\tilde{\gamma}_{\vec{G}_{\alpha}}$. Eq.~\eqref{eq:NewtonRaphDiag} is then restricted to $N_G$ components with the $N_G\times N_G$ diagonal matrix $\left\{\mathbb{J}_D\right\}_{\alpha,\alpha}=\frac{\p \sigma_{\vec{G}_\alpha}}{\p \tilde{\gamma}_{\vec{G}_\beta}}\delta_{\alpha,\beta}$.
Previous experience shows that the use of such diagonal approximation of the Jacobian does not hinder the speed of convergence of the Newton-Raphson process, in particular when --- as usual in applications of SHAKE in classical MD --- this matrix is diagonally dominant. Furthermore, in typical classical MD applications, the evaluation of the diagonal elements of the Jacobian does not pose a numerical challenge because the constraints have simple forms (e.g. fixed bond distances in molecules), and involve small subsets of degrees of freedom (e.g. neighboring atoms in the molecules). This is, unfortunately, not the case for the constraints considered in this work. 
To appreciate the effort required to compute the matrix elements of the Jacobian,  let us consider the different terms in Eq.~\eqref{eq:dsigma_dgamma}.  From the definition of the constraints, we have, for example,
\begin{equation}
\begin{aligned}
 \frac{\p \sigma_{\vec{G}_{\alpha}}(\tilde{\vec{n}}^{\kappa})}{\p \tilde{n}_{\vec{G}_{\lambda}}} &= \frac{\p^2 E[\tilde{\vec{n}}^{\kappa}]}{\p \tilde{n}_{\vec{G}_{\lambda}}\p \tilde{n}\cc_{\vec{G}_{\alpha}}}; \\
 \frac{\p \sigma_{\vec{G}_{\alpha}}(\tilde{\vec{n}}^{\kappa})}{\p \tilde{n}\cc_{\vec{G}_{\lambda}}} &= \frac{\p^2 E[\tilde{\vec{n}}^{\kappa}]}{\p \tilde{n}\cc_{\vec{G}_{\lambda}}\p \tilde{n}\cc_{\vec{G}_{\alpha}}},
\end{aligned}
\end{equation}
with analogous expressions for the other terms. The second derivatives of the energy with respect to the expansion coefficients, in particular for the kinetic and exchange-correlation terms, are typically available via libraries of energy functionals for DFT calculations. In the following, we refer to the \libxc{}~\cite{lehtola:2018} library that has been used in our implementation of MaZe OFDFT. Explicit expressions for the second derivatives are more easily discussed by considering separately the terms in Eq.~\eqref{eq:EnergyODFT}. Let us begin by observing that the Coulomb interaction between the ions (that does not depend on the electronic density) and the interaction energy between electrons and ions, $E_{\rm{ext}}[\vec{R},n]$ (that depends linearly on the electronic density) do not contribute to the second derivatives. The Hartree term, on the other hand, is quadratic in the electronic density. Using its expression in Fourier space, it is immediate to show that
\begin{equation}\label{eq:HartreeHessian}
    \frac{\p ^2 E_{\rm{Hart}}[n]}{\p \tilde{n}_{\vec{G}_{\beta}}\p \tilde{n}\cc_{\vec{G}_{\alpha}}}=\frac{4\pi\Omega}{|\vec{G}_{\alpha}|^2}\delta_{\alpha,\beta},
\end{equation}
indicating that this term is diagonal in reciprocal space.
The evaluation of the second derivatives of the exchange-correlation (xc) and kinetic energy functionals is less straightforward. In the following, we shall consider the GGA class of approximation for these quantities, so they will depend, in general, on the electronic density and its gradient. GGA xc functionals are often used in condensed-phase applications both in orbital-free and Kohn-Sham DFT,\cite{perdew:1992} indicating that the algorithm discussed here has a broad interest. Within OFDFT, GGA kinetic energy functionals include the Thomas-Fermi-von Weizs\"acker~\cite{ vonweizsacker:1935} (TFvW) and Perrot~\cite{perrot:1994} approximations discussed in the Results section. The \libxc{} library processes the kinetic and xc functionals in the same way, and in the following we shall discuss directly the sum of these two terms, i.e. $E_{\text{KXC}}[n,\nabla n] =  E_{\text{KE}}[n,\nabla n] + E_{\text{xc}}[n, \nabla n]$. A description of the manipulations leading from the \libxc{} output to a form of the second derivatives suitable for our purposes is given in Appendix~\ref{app:MV_GGA}. Here we report the final result (see eq.~\eqref{eq:sigma_kxc_app})
\begin{equation}
\label{eq:constr_gradient_text}
\begin{aligned}
&\frac{\p^2 E_{\text{KXC}}}{\p \tilde{n}_{\vec{G}_{\beta}}\p \tilde{n}\cc_{\vec{G}_{\alpha}}} = \frac{\Omega}{N_D}\sum_{\vec{r}}\exp[\ii\vec{G}_{\beta}\cdot\vec{r}]\exp[-\ii\vec{G}_{\alpha}\cdot\vec{r}] \\
&\times \biggl[\Xi(\vec{r}) + \ii(G^l_{\beta} - G^l_{\alpha})\Phi^l(\vec{r})+G^l_{\beta}G^m_{\alpha}\Psi^{lm}(\vec{r})\biggr],
\end{aligned}
\end{equation}
where $N_D$ is the number of points used to discretize the density in real space, $l$ and $m$ identify Cartesian components of the three-dimensional vectors and repetition of these indexes indicates a sum over them.
The functions $\Xi(\vec{r})$, $\Phi^l(\vec{r})$ and $\Psi^{lm}(\vec{r})$ are defined in Appendix~\ref{app:MV_GGA}. Here we note for future convenience that they are real periodic functions in real space.
The second derivatives with respect to different combination of coefficients and complex conjugates of coefficients that are needed in the Jacobian matrix can be obtained from the expression above using the rules of complex conjugation. Note the Hessian matrix above is dominated by its diagonal for which $\vec{G}_{\beta}-\vec{G}_{\alpha}=\vec{0}$ and all terms in the sum over $\vec{r}$ interfere constructively. The same is true for the term $\frac{\p^2 E_{\text{KXC}}}{\p \tilde{n}\cc_{\vec{G}_{\beta}}\p \tilde{n}_{\vec{G}_{\alpha}}}$. However, the two terms $\frac{\p^2 E_{\text{KXC}}}{\p \tilde{n}_{\vec{G}_{\beta}}\p \tilde{n}_{\vec{G}_{\alpha}}}$ and $\frac{\p^2 E_{\text{KXC}}}{\p \tilde{n}\cc_{\vec{G}_{\beta}}\p \tilde{n}\cc_{\vec{G}_{\alpha}}}$, entering in the expressions of $\frac{\p \sigma\cc_{\vec{G}_{\alpha}}}{\p \tilde{n}_{\vec{G}_{\beta}}}$ and  $\frac{\p \sigma_{\vec{G}_{\alpha}}}{\p \tilde{n}\cc_{\vec{G}_{\beta}}}$, respectively, are smaller in magnitude since they all carry a phase with wavevector $\pm(\vec{G}_{\beta}+\vec{G}_{\alpha})$.

The structure of the Hartree and KXC terms motivates our choice of the $2N_G$ complex vectors defined in Eqs.~\eqref{eq:ConstraintVec} and~\eqref{eq:GammaVec}, and to solve the constraint in Fourier space. This choice, in fact, ensures that the Jacobian matrix is diagonally dominant facilitating the convergence of the standard SHAKE algorithm.

Substituting Eq.~\eqref{eq:HartreeHessian} and the second derivatives of the KXC term, Eq.~\eqref{eq:constr_gradient_text}, in Eq.~\eqref{eq:dsigma_dgamma}, we obtain
\begin{widetext}
\begin{equation}\label{eq:SHAKEDenPrestigeStart}
\begin{aligned}
\left[\mathbb{J}_D\right ]_{{\vec{G}_{\alpha}},\vec{G}_{\alpha}}   =&\left(\frac{4\pi\Omega}{|\vec{G}_{\alpha}|^2}\right)^2+ \frac{4\pi\Omega}{|\vec{G}_{\alpha}|^2} \frac{\Omega}{N_D} \sum_{\vec{r}}\Bigl[\Xi_{\text{P}}(\vec{r}) + G^l_{\alpha}G^m_{\alpha}\Psi_{\text{P}}^{lm}(\vec{r})\Bigr]+ \frac{4\pi\Omega}{|\vec{G}_{\alpha}|^2} \frac{\Omega}{N_D} \sum_{\vec{r}}\Bigl[\Xi_{t}(\vec{r}) + G^l_{\alpha}G^m_{\alpha}\Psi_{t}^{lm}(\vec{r})\Bigr]\\
&+ \biggl(\frac{\Omega}{N_D}\biggr)^2\sum_{\vec{r},\vec{r}'}e^{-\ii\vec{G}_{\alpha}\cdot\vec{r}}e^{\ii\vec{G}_{\alpha}\cdot\vec{r}'}\sum^{N_G}_{\lambda=1}e^{\ii\vec{G}_{\lambda}\cdot\vec{r}}e^{-\ii\vec{G}_{\lambda}\cdot\vec{r}'} \\
&\qquad\times \Bigl[\Xi_{\text{P}}(\vec{r}) + \ii (G^l_{\lambda} - G^l_{\alpha})\Phi^{l}_{\text{P}}(\vec{r}) + G^l_{\lambda}G^m_{\alpha}\Psi_{\text{P}}^{lm}(\vec{r})\Bigr]
\Bigl[\Xi_{t}(\vec{r}') - \ii (G^l_{\lambda} - G^l_{\alpha})\Phi_{t}^l(\vec{r}' ) + G^l_{\lambda}G^m_{\alpha}\Psi_{t}^{lm}(\vec{r}')\Bigr] \\
&+ 
\biggl(\frac{\Omega}{N_D}\biggr)^2\sum_{\vec{r},\vec{r}'}e^{-\ii\vec{G}_{\alpha}\cdot\vec{r}}e^{\ii\vec{G}_{\alpha}\cdot\vec{r}'}\sum^{N_G}_{\lambda=1}e^{-\ii\vec{G}_{\lambda}\cdot\vec{r}}e^{\ii\vec{G}_{\lambda}\cdot\vec{r}'} \\
&\qquad\times \Bigl[\Xi_{\text{P}}(\vec{r}) - \ii (G^l_{\lambda} + G^l_{\alpha})\Phi_{\text{P}}^l(\vec{r}) - G^l_{\lambda}G^m_{\alpha}\Psi_{\text{P}}^{lm}(\vec{r})\Bigr]
\times \Bigl[\Xi_{t}(\vec{r}') + \ii (G^l_{\lambda} + G^l_{\alpha})\Phi_{t}^l(\vec{r}' ) - G^l_{\lambda}G^m_{\alpha}\Psi_{t}^{lm}(\vec{r}')\Bigr],
\end{aligned}
\end{equation}
\end{widetext}
where the first term in the RHS arises from the Hartree contribution to the energy functional, while the others stem from the kinetic and xc functionals or mixed terms. 
The Hartree contribution and the mixed terms can be easily seen to scale linearly with respect to the number of grid points. The difficult contributions in terms of numerical complexity are the last two sums, which we now discuss.

The three sums appearing in these kinetic and xc terms indicate that the brute force calculation of these contributions scales cubically with the number of grid points in direct and reciprocal space, a prohibitive bottleneck for realistic calculations. In Appendix~\ref{app:SM_GGA}, we show how this cost can be reduced, for GGA functionals, to $\Omega \log\Omega$ scaling by exploiting periodicity in direct space. This result will be used for comparison purposes in the Results section.

\subsection{SHAKE with linear conjugate gradient (LCG-SHAKE)}\label{subsec:ElberSHAKE}
An alternative way of performing the Newton-Raphson step, Eq.~(\ref{eq:NewtonRaph}), can be derived by building on an approach first suggested by Weinbach and Elber.\cite{weinbach:2005} These authors proposed to obtain the value of the constraints at each iteration $\kappa$ by considering an approximate and, importantly, symmetric form of the Jacobian matrix which enables to solve the Newton-Raphson system using linear conjugate gradient. This algorithm was conceived to facilitate parallelization of classical constrained MD and applied to the simulation of biological systems.\cite{weinbach:2005} In the following we show how the approach can be adapted to the solution of the Newton-Raphson step for MaZe OFDFT and demonstrate that it overcomes the limitations of standard SHAKE by preserving linear scaling for general forms of the energy functional, including those that employ nonlocal kinetic energies.

Let us consider again Eq.~\eqref{eq:NewtonRaph}: defining $\delta \tilde{\vec{\gamma}}=\tilde{\vec{\gamma}}^{\kappa+1}-\tilde{\vec{\gamma}}^{\kappa}$, this equation can be trivially rewritten as the system of linear equations
\begin{equation}\label{eq:NewtonRaphsonDirect}
  \mathbb{J}(\tilde{\vec{n}}^{\kappa}) \delta \tilde{\vec{\gamma}}= - \omega\vec{\sigma}\bigl(\vec{R}(t+\dt),\tilde{\vec{n}}^{\kappa}\bigr). 
\end{equation}
Our goal is to solve this system as efficiently as possible using linear conjugate gradient. In analogy with Ref.~\citenum{weinbach:2005}, that considered the case of a system of real equations, we observe that linear conjugate gradient cannot be immediately applied because the Jacobian matrix is not hermitian (symmetric in Ref.~\citenum{weinbach:2005}) due to the fact that the explicit expressions for its matrix elements (see for example Eq.~\eqref{eq:dsigma_dgamma}) contain derivatives of the constraints evaluated with the density computed at time $t$ and at the provisional value $\tilde{\vec{n}}^{\kappa}$. Similar to Ref.~\citenum{weinbach:2005}, we then approximate the matrix elements of the Jacobian as
\begin{equation}   \label{eq:dsigma_dgammaccsym}
    \begin{aligned}
        \{\mathbb{J}(t)\}_{\alpha,\beta}&=\frac{\p \sigma_{\vec{G}_{\alpha}}}{\p \tilde{\gamma}_{\vec{G}_{\beta}}} \\ &\approx 
        \sum^{N_G}_{\lambda=1}\biggl[\frac{\p \sigma_{\vec{G}_{\alpha}}(t)}{\p \tilde{n}_{\vec{G}_{\lambda}}}\frac{\p \sigma\cc_{\vec{G}_{\beta}}(t)}{\p \tilde{n}\cc_{\vec{G}_{\lambda}}} + \frac{\p \sigma_{\vec{G}_{\alpha}}(t)}{\p \tilde{n}\cc_{\vec{G}_{\lambda}}}\frac{\p \sigma\cc_{\vec{G}_{\beta}}(t)}{\p \tilde{n}_{\vec{G}_{\lambda}}} \biggr ]   \\
        &= \left\{ \Sigma(t)\cdot \Sigma(t)^\dagger \right\}_{\alpha,\beta},
        \end{aligned}
\end{equation}
and similarly for the other blocks of the matrix.
Compared to the exact SHAKE expression, all derivatives in the equations above are evaluated using the density at time $t$, with similar expressions for the other blocks in Eq.~\eqref{eq:Jacobian}. We denote this approximate Jacobian as $\mathbb{J}\bigl(\tilde{\vec{n}}(t)\bigr)\equiv \mathbb{J}(t)$. Direct inspection of the approximate Jacobian shows that it is hermitian. 
Furthermore, we have
\begin{equation}
    {\vec{\zeta}}\cc\mathbb{J}(t){\vec{\zeta}}= {\vec{\zeta}}\cc\Sigma(t)\Sigma^\dagger(t){\vec{\zeta}}>0,
\end{equation}
for all ${\vec{\zeta}}\neq\vec{0}$, proving that the matrix is positive definite. The solution of 
\begin{equation}\label{eq:NewtonRaphsonElber}
  \mathbb{J}(t) \delta \tilde{\vec{\gamma}}= - \omega\vec{\sigma}\bigl(\vec{R}(t+\dt),\tilde{\vec{n}}^{\kappa}\bigr), 
\end{equation}
(that we take as an approximant for Eq.~\eqref{eq:NewtonRaphsonDirect}) can then be obtained by direct application of linear conjugate gradient with complex variables, with or without preconditioning.\cite{joly:1993} (Note that using linear conjugate gradient, requires the application of the matrix $\mathbb{J}(t)$ on complex vectors of size $2N_G$. This operation can be performed with $\Omega \log \Omega$ scaling for either GGA (see Appendix~\ref{app:MV_GGA}) and nonlocal (see Appendix~\ref{app:MV_nonlocal}) energy functionals.) In Ref.~\citenum{weinbach:2005} the authors showed that the solution of this system converges to the same set of Lagrange multipliers of the standard SHAKE. In the Results section, we also show that this method is more efficient than standard SHAKE when applied to OFDFT dynamics. 

\section{Benchmarks and Results}
\label{sec:results}
The algorithms described in the previous sections are tested and compared in the orbital-free DFT framework on constant energy (NVE) and constant temperature (NVT) simulations of a system of particles of sodium in different thermodynamic states and with different kinetic energy functionals. We start by detailing the relative efficiency of the two methods and then, having shown its merits, we move to compute static and dynamical statistical properties of the system with the LCG-SHAKE. In particular, the reliability of the algorithm for NVT calculations (never tested before), is demonstrated. Finally, we discuss preliminary comparisons the performance of MaZe to that of standard nonlinear conjugate gradient (NLCG) based Born-Oppenheimer dynamics by considering energy conservation and time reversibility for a realistic test system. 

For all calculations discussed in this section, periodic boundary conditions are enforced in all directions and the electrostatic interactions are treated via the Ewald summation method with a cutoff radius of $R_{\text{c}} = 1.06\um{\angstrom}$. For these tests, Slater~\cite{dirac:1930,bloch:1929} and Perdew \& Zunger~\cite{perdew:1981} functionals are used for the exchange and correlation contributions $E_{\text{xc}}$ to the total energy, respectively. Functionals adopted for the kinetic energy are detailed in the following. The interaction of the atomic nucleus with the electrons up to the last closed shell is modeled using the pseudopotential for sodium described by Topp and Hopfield.~\cite{topp:1973} The cutoff for the plane-wave expansion is set to $G_{\text{c}} = 10\um{\hartree}$ and the timestep is set to $\delta t = 1\um{fs}$ for all simulations. The NVE evolution of the physical degrees of freedom is performed using the velocity Verlet algorithm. As for the evolution of the coefficients of the density, the provisional values used as the initial guess for the Newton–Raphson iterative procedure in SHAKE is computed through the Verlet algorithm. The first two values of the coefficients --- necessary for the initialization of the Verlet algorithm --- are obtained through two minimizations of the energy functional using a NLCG algorithm and the full MaZe dynamics for the system is started from the second simulation step. The NVT ensemble is sampled via the use of a Langevin thermostat where the evolution of the physical degrees of freedom is integrated using the algorithm proposed by Vanden-Eijnden and Ciccotti.\cite{vanden-eijnden:2006} The same approach is also used in the NVE runs --- with the friction parameter of the thermostat set to $\nu = 10^{-5}\um{au}$ --- to equilibrate the systems to the target temperatures (see below), via runs of duration equal to $10\um{ps}$. The number of electrons is kept constant along the simulation by excluding the $\vec{G} = (0,0,0)$ component of the electronic density expansion in reciprocal space from the dynamical system as detailed in the previous sections. Tests are focused on two different thermodynamic states of the system,\footnote{The melting temperature of Sodium at room pressure is $T_m = 371\um{K}$.} solid ($T_{s}\approx25\um{K}$, $\rho_s = 969.5\um{kg}\um{m^{-3}}$) and liquid ($T_{l}\approx430\um{K}$, $\rho_l = 926.8\um{kg}\um{m^{-3}}$). 

\begin{figure*}[htb]
\begin{center}
\includegraphics[width=\textwidth]{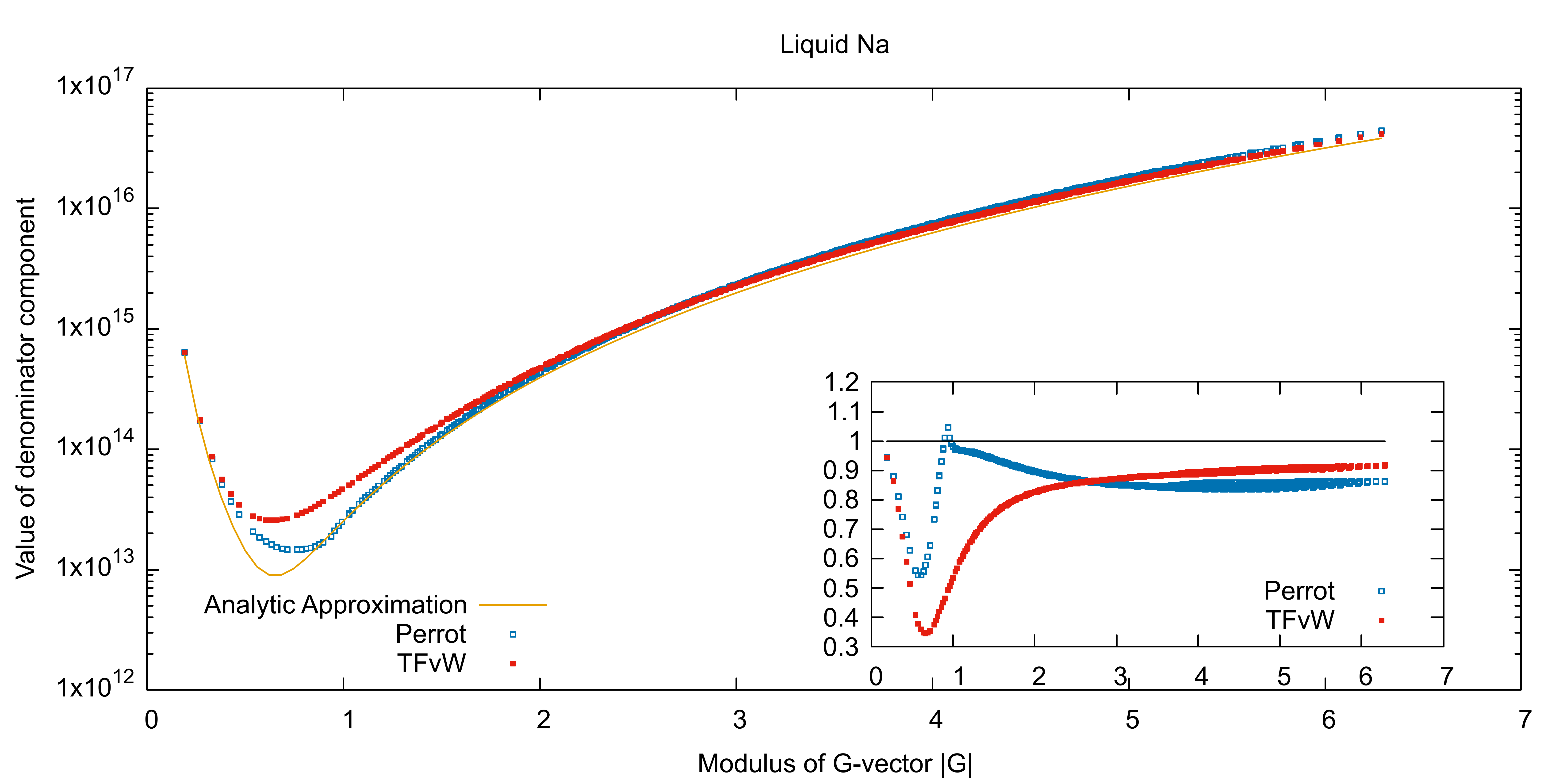}
\end{center}
\caption{Main plot: Diagonal of the SHAKE matrix (SHAKE denominator) for liquid sodium for different kinetic energy functionals (red points TFvW, blue points Perrot) compared with the analytic approximation obtained from Eq.~\eqref{eq:shake_denom_analytic} (yellow curve). Inset: Ratio between the SHAKE denominator and the analytic approximation.} 
\label{fig:liquid_denom}
\end{figure*}

To set the stage for the production runs, we recall that, as discussed in Sec.~\ref{subsec:StandardSHAKE}, when using the standard SHAKE algorithm, the calculation of the matrix $\mathbb{J}_D$ defined in Eq.~\eqref{eq:diagonal} is the most expensive ingredient in the solution of the constraints. To explore this calculation, we implemented and computed it, as per its $\Omega \log\Omega$ scaling expression in Eq.~\eqref{eq:theprestige}, for a system of $N=128$ sodium atoms in the solid (box size of $L_s = 17.1\um{\angstrom}$) and liquid (box size $L_l = 17.4\um{\angstrom}$). We considered two different kinetic energy functionals: Thomas-Fermi-von Weizs\"acker\cite{vonweizsacker:1935} (TFvW) and Perrot.\cite{perrot:1994} Two simulations of 1000 steps for the solid and liquid systems were performed, computing $\mathbb{J}_D$ for configurations obtained every 100 steps. Results for liquid Sodium are reported as a function of the modulus of the wave-vector, as red (TFvW) and blue (Perrot) squares in Figure~\ref{fig:liquid_denom}. Calculations for the solid (not reported) show very similar behavior. Note that the results for the 10 configurations selected in these test runs are essentially superimposed for both curves. These plots show then that the shape of the denominator does not change significantly along the simulation. This behavior is an additional confirmation of the fact that the Hessian of the Hartree term in the energy functional --- independent of the electronic density and thus constant along the simulation --- has a dominant role in the computation of the denominator, therefore corroborating the discussion after Eq~\eqref{eq:HartreeHessian} and Eq.~\eqref{eq:constr_gradient_text}. 

In Figure~\ref{fig:liquid_denom}, the numerical values of the SHAKE denominator are also compared to the analytical approximation:
\begin{equation}
\label{eq:shake_denom_analytic}
\left[\mathbb{J}_D\right ]^{\text{analytical}}_{{\vec{G}_{\alpha}},\vec{G}_{\alpha}} = \biggl(\frac{\Omega}{4\rho}\biggr)^2\vert \vec{G}_\alpha \vert^4 + \bigl(4\pi\Omega\bigr)^2\vert \vec{G}_\alpha \vert^{-4},
\end{equation}
where $\Omega$ is the volume of the simulation box and $\rho = \Ne/\Omega$ is the uniform electron density of the system. This analytical approximation is obtained as the sum of the diagonal Hartree contribution, which is dominant at small $\vert \vec{G}_\alpha \vert$, and of the von~Weizs\"acker contribution to the kinetic energy functional, dominant at large $\vert \vec{G}_\alpha \vert$. The latter is given by
\begin{equation}
    E_{\mathrm{vW}}[n]=\frac{1}{8}\int_\Omega \dd^3\vec{r} \frac{\vert \nabla n(\vec{r})\vert^2}{n(\vec{r})}.
\end{equation}
Assuming $n(\vec{r})=\rho$ constant at the denominator in the integral, we have
\begin{equation}
\begin{aligned}
    E_{\mathrm{vW}}[n] \approx \frac{1}{8\rho}\int_\Omega \dd^3\vec{r} \vert \nabla n(\vec{r})\vert^2\approx \frac{\Omega}{4\rho}\sum_{\alpha=1}^{N_G} \vert \vec{G}_\alpha \vert^2 
     \tilde{n}\cc_{\vec{G}_\alpha} \tilde{n}_{\vec{G}_\alpha}.
\end{aligned}
\end{equation}

In Figure~\ref{fig:liquid_denom} the analytical approximation, Eq.~\eqref{eq:shake_denom_analytic}, is shown as the yellow curve and turns out to be remarkably accurate, at least for the system under investigation. (Similar results were obtained for solid Sodium, not shown.) Adoption of this expression for the SHAKE denominator might be used to reduce the computational cost of our calculations. In the following, however, we employ it only as an effective preconditioner in the LCG-SHAKE algorithm described in Section~\ref{subsec:ElberSHAKE}. 

\begin{figure*}[htb]
\begin{center}
\includegraphics[width=\textwidth]{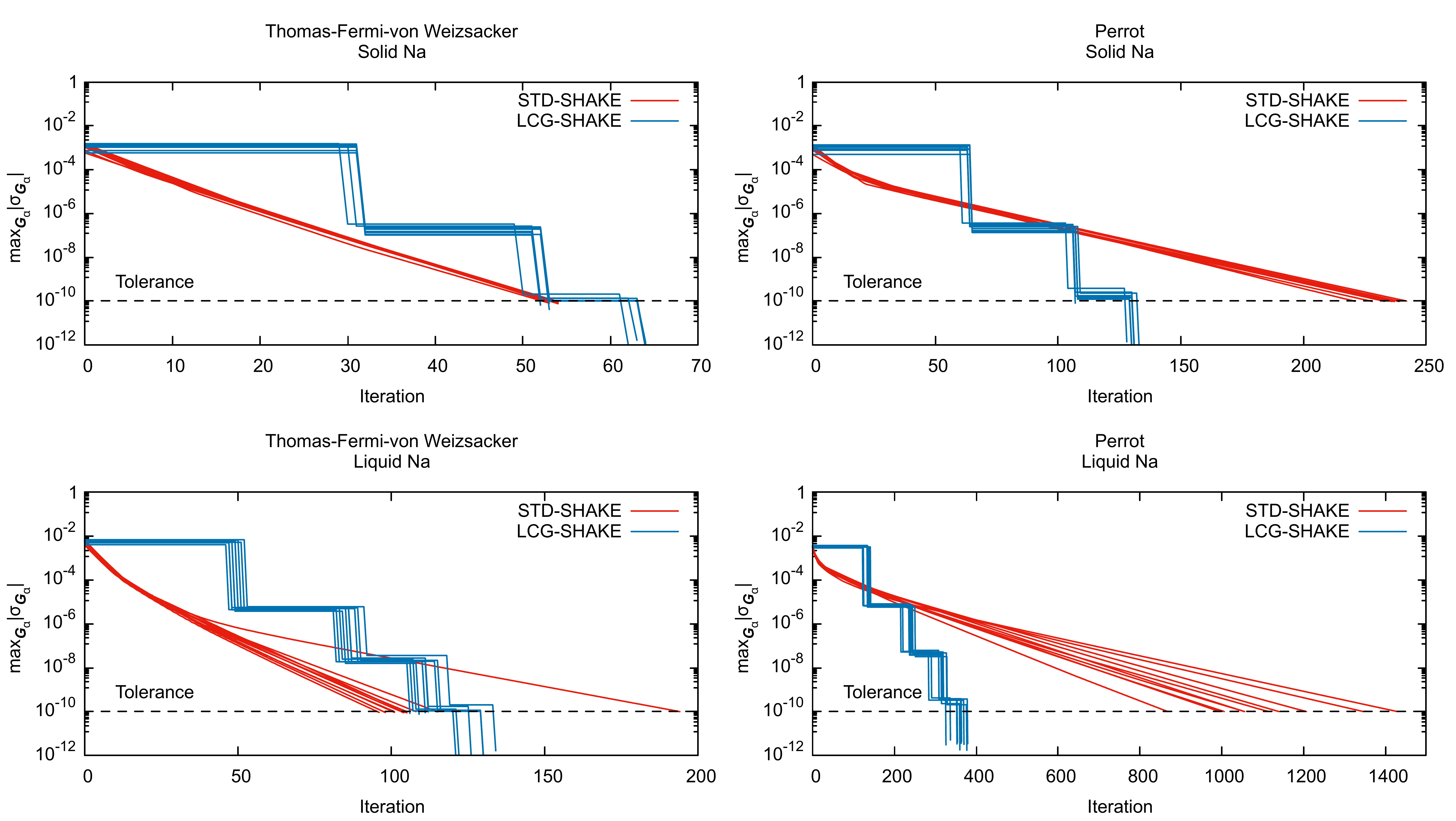}
\end{center}
\caption{Total number of iterations to reach convergence for Newton-Raphson solution of the equation of constraints via standard SHAKE and LCG-SHAKE algorithms. In the LCG-SHAKE case the horizontal (parallel to the $x$-axis) segments represent the number of linear CG iterations used to solve the linearized equation of the constraints, as the values on the $y$-axis refers to the Newton-Raphson iterations.} 
\label{fig:convergence_comparison}
\end{figure*}

\begin{figure}[htb]
\begin{center}
\includegraphics[width=\columnwidth]{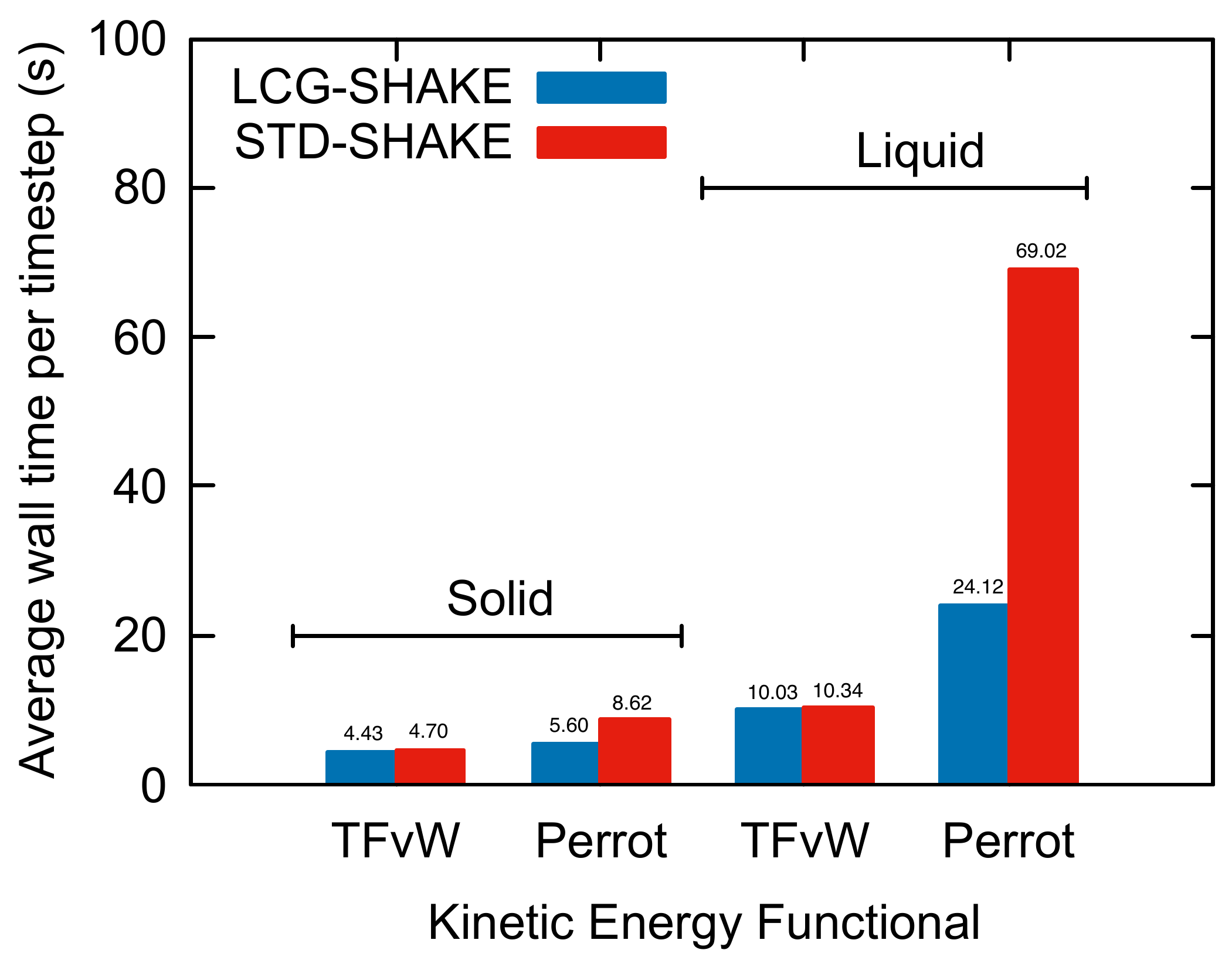}
\end{center}
\caption{Average wall time per timestep comparison between Standard SHAKE and LCG-SHAKE algorithms for different kinetic energy functionals and different thermodynamic states.} 
\label{fig:time_comparison}
\end{figure}

We now move to the comparison of the efficiency of the standard SHAKE and LCG-SHAKE implementations of MaZe. The tests below have been performed using the same set of configurations employed in the discussion above. We begin by showing, in Figure~\ref{fig:convergence_comparison}, the path to convergence of the Newton-Raphson solution of the constraint equations for the two different algorithms. The figure reports, as a function of the total number of iterations needed to converge, the modulus of the largest component of the constraint vector below a given threshold. In red, results for the standard SHAKE are reported, showing the characteristic monotonic decrease of $\mathrm{max}_{\vec{G}_{\alpha}}|\sigma_{\vec{G}_{\alpha}}|$ with number of iterations already observed in previous MaZe implementations.\cite{bonella:2020} The blue curves, on the other hand, show the path to convergence of the LCG-SHAKE. We recall that for the latter, the linear system, Eq.~\eqref{eq:NewtonRaphsonDirect}, needs to be solved via linear conjugate gradient, implying an additional iterative procedure. This linear system is considered solved when the square-root of the norm of the residuals is less than $5\cdot10^{-14}\um{au}$. The iterations needed for this solution are represented in Figure~\ref{fig:convergence_comparison} as the horizontal segments. Both for standard SHAKE and for the LCG-SHAKE, we set the convergence threshold of the Newton-Raphson algorithm to $10^{-10}\um{au}$.  A successive over-relaxation parameter $\omega = 0.35$ and $\omega = 0.21$ is set in the standard SHAKE algorithm when the TFvW and Perrot kinetic energy functionals are used, respectively. For the LCG-SHAKE algorithm, $\omega = 1$ proved the best choice in all tests performed. The panels in the figure show that, for the functional considered and irrespective of the solid (upper panel) or liquid (lower) state of the system, both algorithms converge in a relatively small number of iterations. Furthermore, LCG-SHAKE clearly outperforms standard SHAKE when the Perrot functional is adopted. The apparent advantage of standard SHAKE for the TFvW functional is however limited to the number of iterations. In fact, comparing the efficiency of the two algorithms in terms of average wall time to solution, Figure~\ref{fig:time_comparison}, we see that the algorithms (both very efficient) have relatively similar performances for the TFvW calculations, while LCG-SHAKE is faster than standard SHAKE for Perrot, most notably in the case of the liquid simulations.

\begin{figure}[htb]
\begin{center}
\includegraphics[width=\columnwidth]{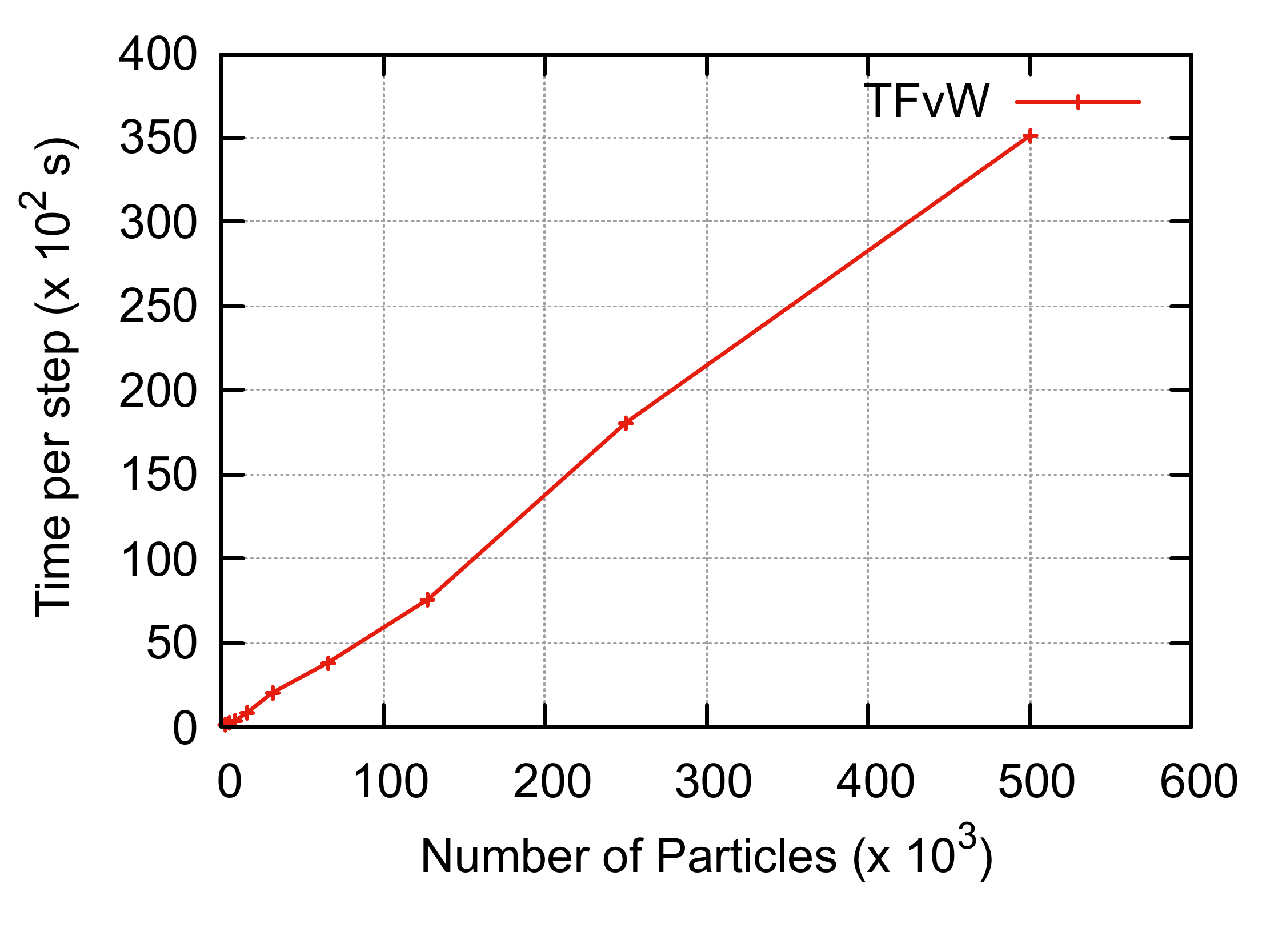}
\end{center}
\caption{Performance scaling of a system of solid sodium with TFvW kinetic energy functional with respect to the number of particles for the LCG-SHAKE algorithm.} 
\label{fig:scaling}
\end{figure}

We conclude this survey of the numerical properties of MaZe by demonstrating, for the more efficient LCG-SHAKE, the linear scaling of the computational cost with respect to the system size. For this particular benchmark, 10 simulation steps of a system of solid sodium modelled with the TFvW kinetic energy functional have been performed for a number of particles ranging from $N=2000$ to $N=500094$, and the average wall time per timestep has been reported in Figure~\ref{fig:scaling} as a function of the number of particles in the simulation, showing a very good linear scaling.

Accuracy and reliability of the LCG-SHAKE MaZe method in performing physically realistic simulations are now shown on a system of liquid sodium modelled through the Perrot kinetic energy functional. Two sets of simulations are presented in the following, in the NVE and NVT ensemble. The temperature of the system is the same for both sets of simulations, $T = 434\um{K}$. 
In all cases, the calculation of the coefficients of the density in reciprocal space is performed using MaZe together with the LCG-SHAKE algorithm, where the provisional value of the additional dofs is computed through a standard Verlet algorithm.

\begin{figure}[htb]
\begin{center}
\includegraphics[width=\columnwidth]{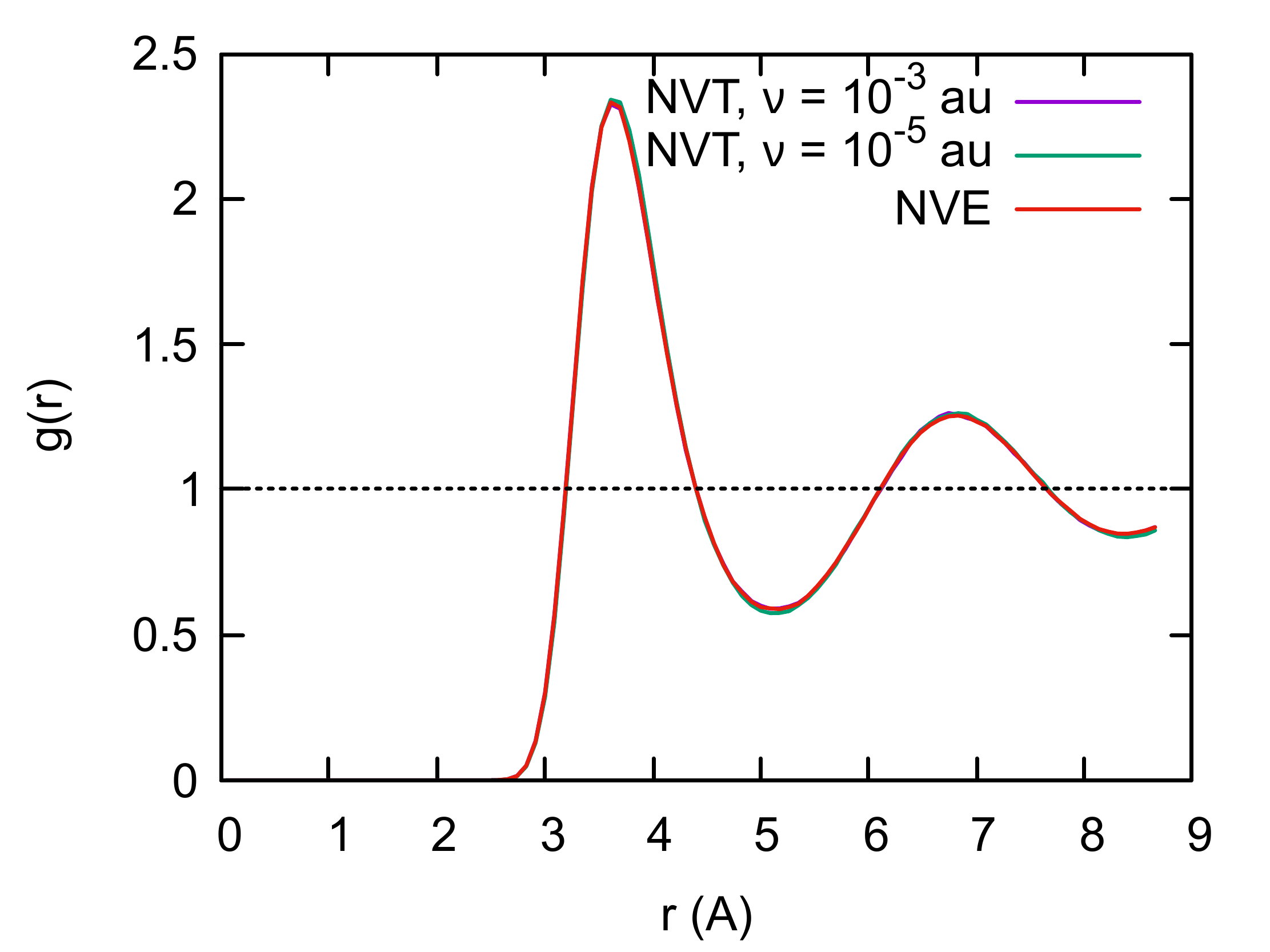}
\end{center}
\caption{Comparison between radial distribution functions from NVT runs performed using the algorithm in Ref.~\citenum{vanden-eijnden:2006} for a Langevin thermostat for different values of the friction parameters $\nu$. The NVE result is obtained from a 100\um{ps} simulation of $N=128$ particles of sodium at $T=434$\um{K}. The other values of $\nu$ are obtained from 50\um{ps} NVT simulations of $N=128$ particles of sodium at $T^* = 434$\um{K}.} 
\label{fig:NVT_GofR}
\end{figure}

\begin{figure}[htbp]
\begin{center}
\includegraphics[width=\columnwidth]{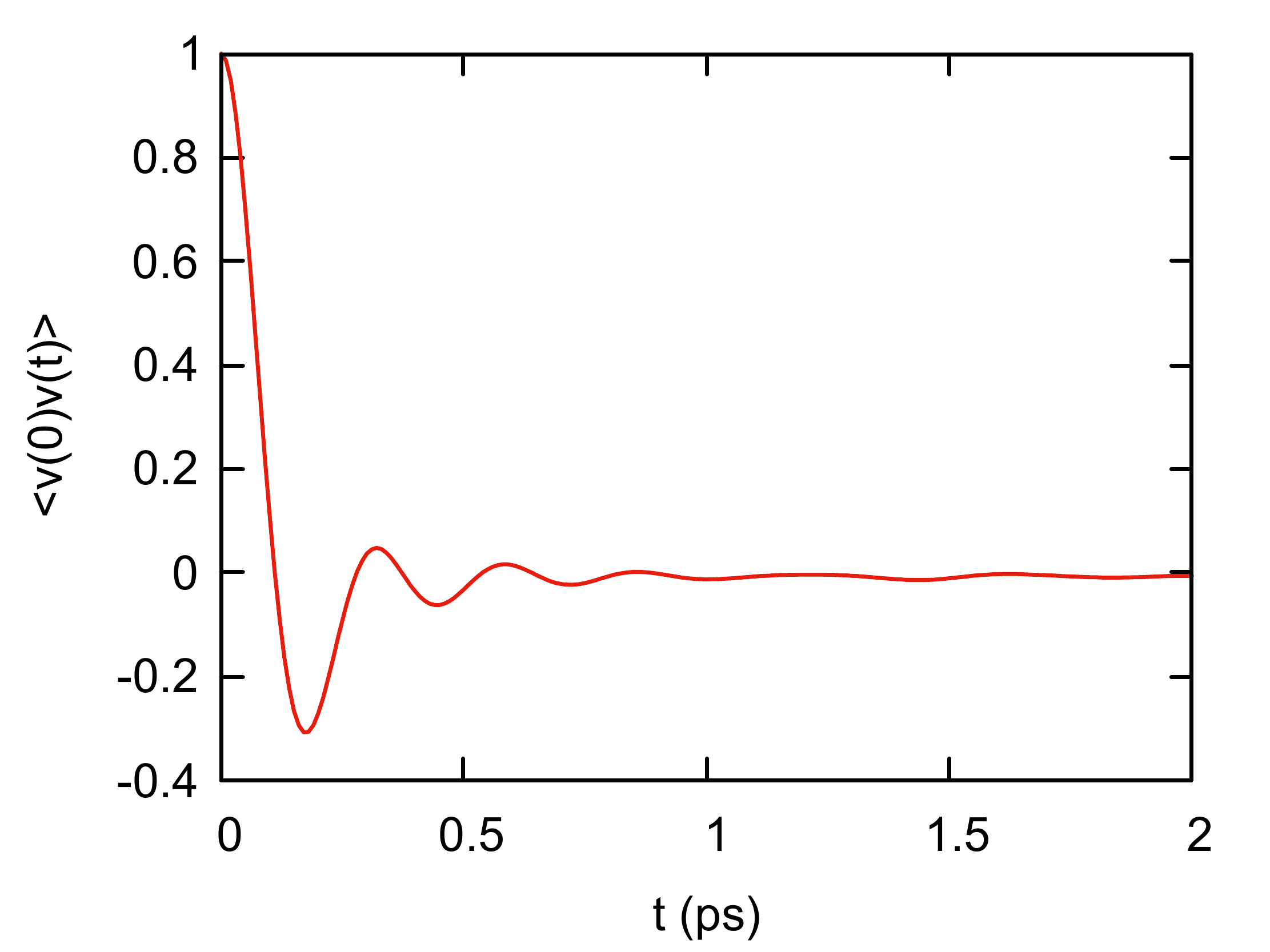}
\end{center}
\caption{Normalized velocity autocorrelation function for a 100\um{ps} NVE simulation of $N=128$ particles of liquid sodium at $T=434$\um{K}.} 
\label{fig:NVE_LRT}
\end{figure}

\begin{figure}[htb]
\begin{center}
\includegraphics[width=\columnwidth]{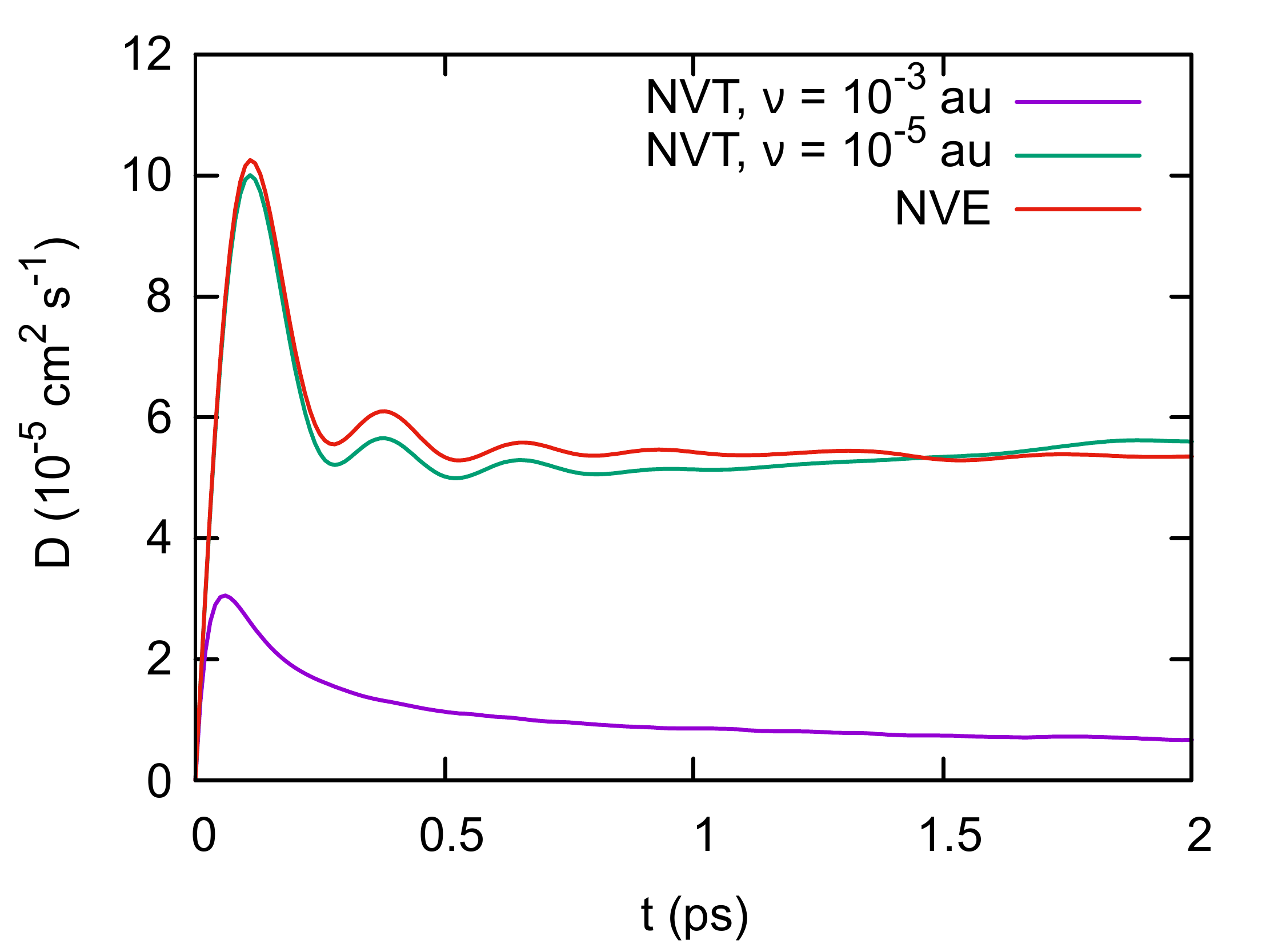}
\end{center}
\caption{Comparison between diffusion coefficients from NVT runs 
for different values of the friction parameter $\nu$. The NVE result is obtained from a 100\um{ps} simulation of $N=128$ particles of Na at $T=434$\um{K}. The other values of $\nu$ are obtained from 50\um{ps} NVT simulations of $N=128$ particles of Na at $T^* = 434$\um{K}.} 
\label{fig:NVT_Diff}
\end{figure}

We begin by considering the radial distribution function of the system, shown in Figure~\ref{fig:NVT_GofR}. The red curve shows the result obtained for an NVE run of duration $100\um{ps}$. Position and height of the two peaks are in agreement with previous numerical studies~\cite{qian:1990} performed at the same temperature and are compatible with experimental results~\cite{greenfield:1971,lee:1978} obtained at $T_{\text{exp}} = 373\um{K}$. The green and purple curves, on the other hand, report NVT results for two different values of the friction coefficient $\nu$ obtained from $50\um{ps}$ long simulations. As expected, within noise, this static property is independent of the choice of the friction and identical to the NVE output. Figure~\ref{fig:NVE_LRT} shows the velocity autocorrelation function obtained from the constant energy run. Both the characteristics of this curve and the diffusion coefficient obtained integrating the result over time (shown as the red curve in Figure~\ref{fig:NVT_Diff}) are in very good agreement with previous numerical results.\cite{shimojo:1994} In particular, our computed value of the NVE diffusion is equal to  $5.37\pm0.16\um{cm^2}\um{s^{-1}}$ which agrees with the numerical results reported in Ref.~\citenum{qian:1990} and with the experimental results reported in Ref.~\citenum{ozelton:1968} for the temperature of $T = 434\um{K}$ at which the simulations are performed. Figure~\ref{fig:NVT_Diff} shows also the value of the diffusion coefficients computed in NVT simulations with different values of the friction coefficients. In this case, the result naturally depends on the choice of $\nu$, but we observe that, again as expected, it approaches the reference NVE for small friction further illustrating the reliability of our canonical MaZe.

\begin{figure}[htb]
\begin{center}
\includegraphics[width=\columnwidth]{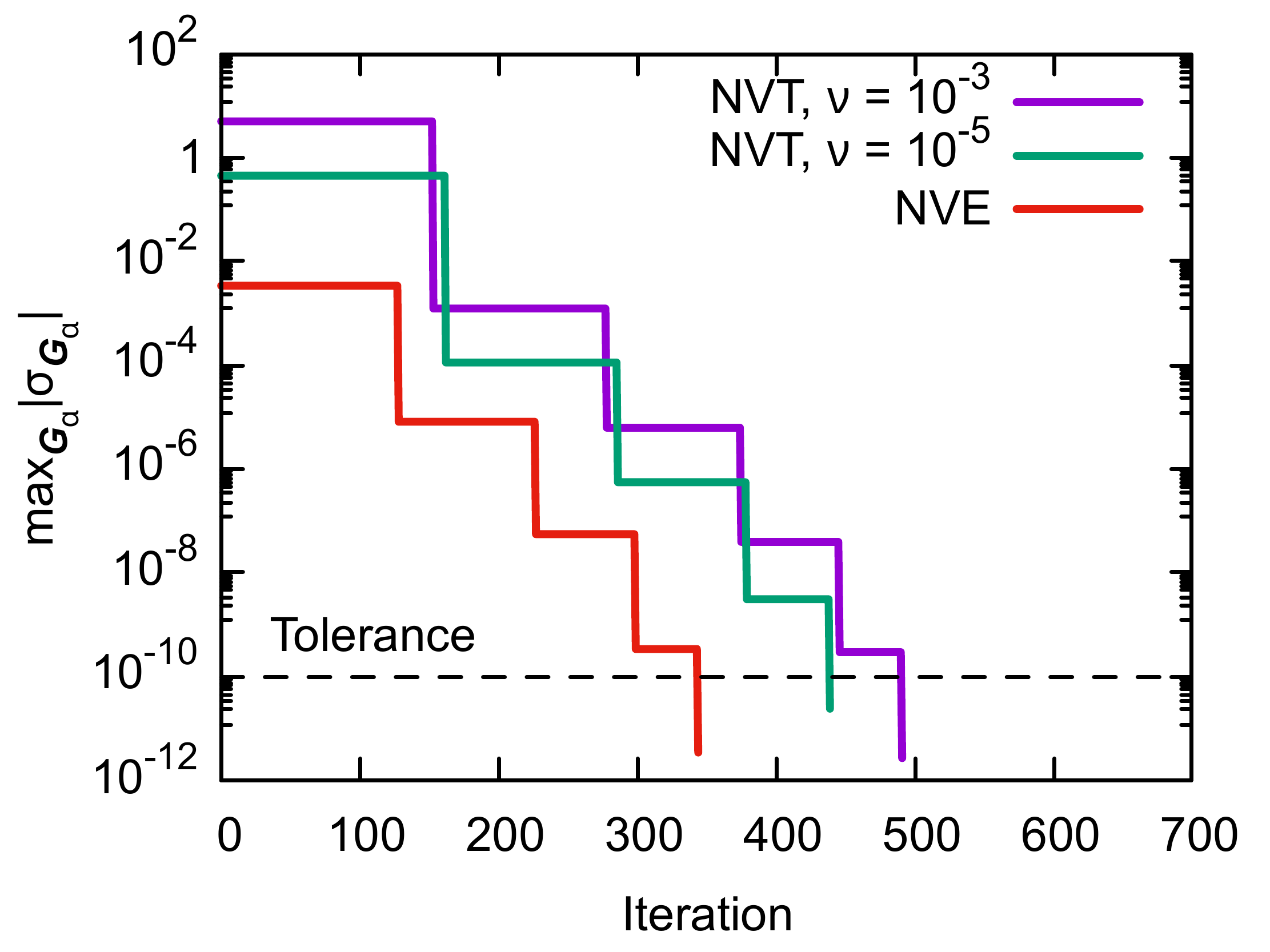}
\end{center}
\caption{Total number of iterations to reach convergence for NVT simulation with LCG-SHAKE for different values of the friction parameter $\nu$. The horizontal (parallel to the $x$-axis) segments represent the number of linear CG iterations used to solve the linearized equation of the constraints, as the values on the $y$-axis refers to the Newton-Raphson iterations. Results are for a typical minimization path along a simulation.} 
\label{fig:convergence_NVT}
\end{figure}

In the NVT ensemble, the number of iterations to converge the calculation of the electronic density is also affected by the choice of $\nu$. In Figure~\ref{fig:convergence_NVT} the path to convergence for 
a typical 
run is shown for two different values of the friction coefficient: $\nu=10^{-3}$\um{au}, in purple, and $\nu=10^{-5}$\um{au}, in green, and compared with the NVE result, in red.
The trends in Figure~\ref{fig:convergence_NVT} indicate that the increase in the number of iterations necessary to solve the equation of the constraints with increasing $\nu$ is mainly due to a difference in the value of the maximum component of the constraints at the start of the iterative loop. Since the thermostat only affects the dynamics of the physical degrees of freedom, this depends on the fact that the provisional value of the electronic coefficients used to initialize the SHAKE routine at each timestep, is a less accurate first guess for higher values of the ionic friction. Indeed, once the initial penalty associated to the violation of the constraint at zero iteration is absorbed, typically in one Newton-Raphson step, the speed of convergence becomes comparable with that of NVE. Furthermore, even for large values of $\nu$, the total number of iterations to convergence remains relatively small. 

We conclude this section by benchmarking LCG-SHAKE MaZe 
against standard Born-Oppenheimer dynamics, based on nonlinear conjugate-gradient (NLCG) minimization.\cite{press:1992-book} 
Tests are particularly focused on the time-reversibility properties of the algorithms and relative efficiency of the two approaches. Note however that for Born-Oppenheimer dynamics we have used a simple textbook implementation of the NLCG algorithm. The initial density for NLCG is taken at each timestep to be the final density of the last timestep, thus no extrapolation was performed. Similarly, no preconditioning of the NLCG was used. The comparison is performed on a set of simulations of $N=16$ particles of liquid sodium ($L=8.68\um{\angstrom}$) using the Perrot kinetic energy functional.\cite{pearson:1993,perrot:1994} The low number of particles has been chosen to ensure that possible numerical errors arising in the simulations are mainly due to the propagation algorithm and not to the computation of the forces. Time reversibility of the algorithms is tested as follows: Starting from the same initial configuration, a $\mathcal{T}=10\um{ps}$ simulation is performed using either nonlinear conjugate-gradient Born-Oppenheimer or MaZe to enforce the minimum energy condition (forward trajectory). The velocities of the final configuration of the forward runs are then reversed and a new 10\um{ps} simulation is performed (backward trajectory). Along the forward and backward trajectories, positions of the particles are recorded every 10\um{fs}. Average root mean square displacements normalized to the length of the simulation box
\begin{equation}
\label{eq:RMSD}
\mathrm{RMSD}(t) = \frac{1}{NL}\sum_{i=1}^{N}\sqrt{\bigl(\vec{r}_i^+(\mathcal{T}-t) - \vec{r}_i^-(t)\bigr)^2},
\end{equation}
are computed for corresponding configurations along backward and forward trajectories for the two algorithms. The convergence threshold of the minimum condition is set to a very stringent value (residual lower than $10^{-10}$\um{au} for the modulus of the maximum component of the derivative of the energy, compare with Tables~\ref{tab:TimeRev_MaZe} and ~\ref{tab:TimeRev_NLCG}).
\begin{figure*}[htb]
\begin{center}
\includegraphics[width=\textwidth]{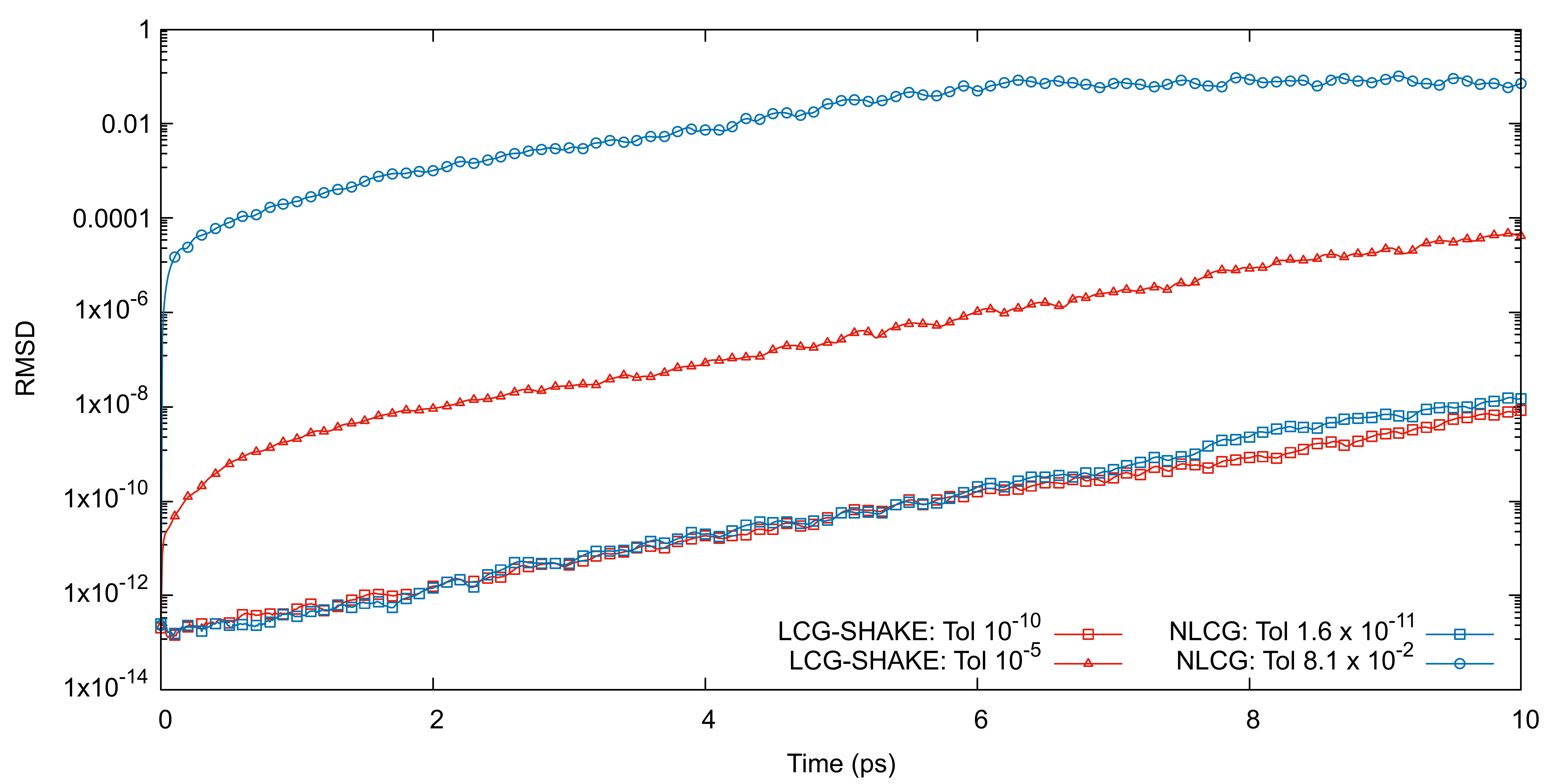}
\end{center}
\caption{Test of time-reversibility properties of MaZe (LCG-SHAKE, curves with red symbols) with respect to standard Born-Oppenheimer alternatives, where the minimization of the energy functional is performed with standard nonlinear conjugate-gradient (NLCG, curves with blue symbols) techniques with different convergence thresholds. The test is based on comparing configurations along a forward- and a backward-in-time trajectory and comparing the average RMSD for corresponding configurations (see Eq.~\eqref{eq:RMSD} and relative discussion). The upper limit on the $y$-axis represents an estimate of the maximum allowed value of the RMSD considering the presence of the PBCs (and therefore the fact that two particles are never more than $\frac{L\sqrt{3}}{2}$ distant from each other). An intermediate threshold of $10^{-5}$ for MaZe is also presented (red line open triangles) to demonstrate the levels of time reversibility and stability attainable with intermediate tolerances.}
\label{fig:TimeRev}
\end{figure*}
Results for LCG-SHAKE MaZe and NLCG Born-Oppenheimer with this threshold are shown as the red and blue open squares symbols, respectively, in Figure~\ref{fig:TimeRev}. The figure shows that results for the two algorithms are very similar with this strict tolerance. The slow violation of time reversibility observed can be attributed to the numerical error associated to the finite timestep and round off that are amplified by the Lyapunov instability. 

With this strong convergence requirement, however, the average time needed by NLCG to solve the minimization problem is ten times larger than the one needed by MaZe to solve the equation of the constraints. Table~\ref{tab:TimeRev_MaZe} reports timings for LCG-SHAKE for different tolerances. For each user-chosen tolerance, we report not only the observed timing but also the actual residual achieved. The final residual can be significantly lower than the target tolerance as only one Newton-Raphson (NR) iteration can lead to high convergence. The average number of NR iterations is also reported in this table.
\begin{table}[hbt]
\caption{Performance assessment for MaZe using LCG-SHAKE algorithm for the solution of the constraint equation. The user-chosen tolerance of the iterative procedure is reported in the second column, while in the third column the actual residual on the constraints after the self-consistent cycle is shown. Wall timings for a single MD step on a single core (AMD Ryzen 7 5800X CPU) and number of Newton-Raphson iterations to reach convergence are reported in the fourth and fifth column respectively.}
\label{tab:TimeRev_MaZe}
\begin{ruledtabular}
\begin{tabular}{lllll}
Method & Tol ($\um{au}$) & Res ($\um{au}$) & Time (s) & NR Iters \\
\hline
\multirow{5}{*}{LCG-SHAKE}& $10^{-1}$ & $1.4\times10^{-4}$ & $3.3\times10^{-2}$ & 1\\
& $10^{-3}$ & $2.0\times10^{-6}$ & $5.8\times10^{-2}$ & 1\\
& $10^{-5}$ & $1.3\times10^{-6}$ & $8.0\times10^{-2}$ & 1\\
& $10^{-7}$ & $5.0\times10^{-9}$ & $1.5\times10^{-1}$ & 2\\
& $10^{-10}$ & $2.2\times10^{-11}$ & $2.7\times10^{-1}$ & 3\\
\end{tabular}
\end{ruledtabular}
\end{table}

For comparison, Table~\ref{tab:TimeRev_NLCG} reports the timings for our textbook implementation of the NLCG method. In order to allow for a one to one correspondence with the LCG-SHAKE case, the NLCG tolerance is set to the observed residual in Table~\ref{tab:TimeRev_MaZe}.
\begin{table}[hbt]
\caption{Performance assessment of standard Born-Oppenheimer MD using nonlinear conjugate gradient algorithm. Comparison between the methods is carried out by setting the same tolerance (second column) for the nonlinear conjugate gradient method as the residual on the constraints shown in Table~\ref{tab:TimeRev_MaZe}. Wall timings for a single MD step for the same system used for tests reported in Table~\ref{tab:TimeRev_MaZe} are reported in the last column. An additional high tolerance is shown in the first line where the performances in terms of wall time are comparable between the two methods. Note that MaZe algorithm is faster than NLCG even when the precision is nine order of magnitudes higher than standard nonlinear conjugate gradient.}
\begin{ruledtabular}
\label{tab:TimeRev_NLCG}
\begin{tabular}{lll}
Method & Tol ($\um{au}$) & Time (s)\\
\hline
\multirow{5}{*}{NLCG}& $8.1\times10^{-2}$ & $3.8\times10^{-1}$ \\
& $8.8\times10^{-5}$ & $9.9\times10^{-1}$ \\
& $1.3\times10^{-6}$ & 2.3 \\
& $4.4\times10^{-9}$ & 3.7 \\
& $1.6\times10^{-11}$ & 4.6\\
\end{tabular}
\end{ruledtabular}
\end{table}
To obtain comparable average times per timestep along the simulation, the tolerance of the nonlinear conjugate gradient has to be relaxed by nine orders of magnitude compared the lowest tolerance achieved with LCG-SHAKE.
NLCG runs at low convergence, however, lead to dramatic effects and, in particular, the algorithm suffers an essentially complete loss of time reversibility as shown in Figure~\ref{fig:TimeRev}, where the blue open circles shows that the average distance between time-reversed configurations quickly reaches the box dimensions.
We also report in Figure~\ref{fig:TimeRev}, the RMSD for an LCG-SHAKE run with intermediate tolerance, corresponding to only one NR iteration per timestep. It can be seen that such a trajectory is stable, with an RMSD that parallels the RMSD with the lowest tolerance.

\begin{figure}[htb]
\begin{center}
\includegraphics[width=\columnwidth]{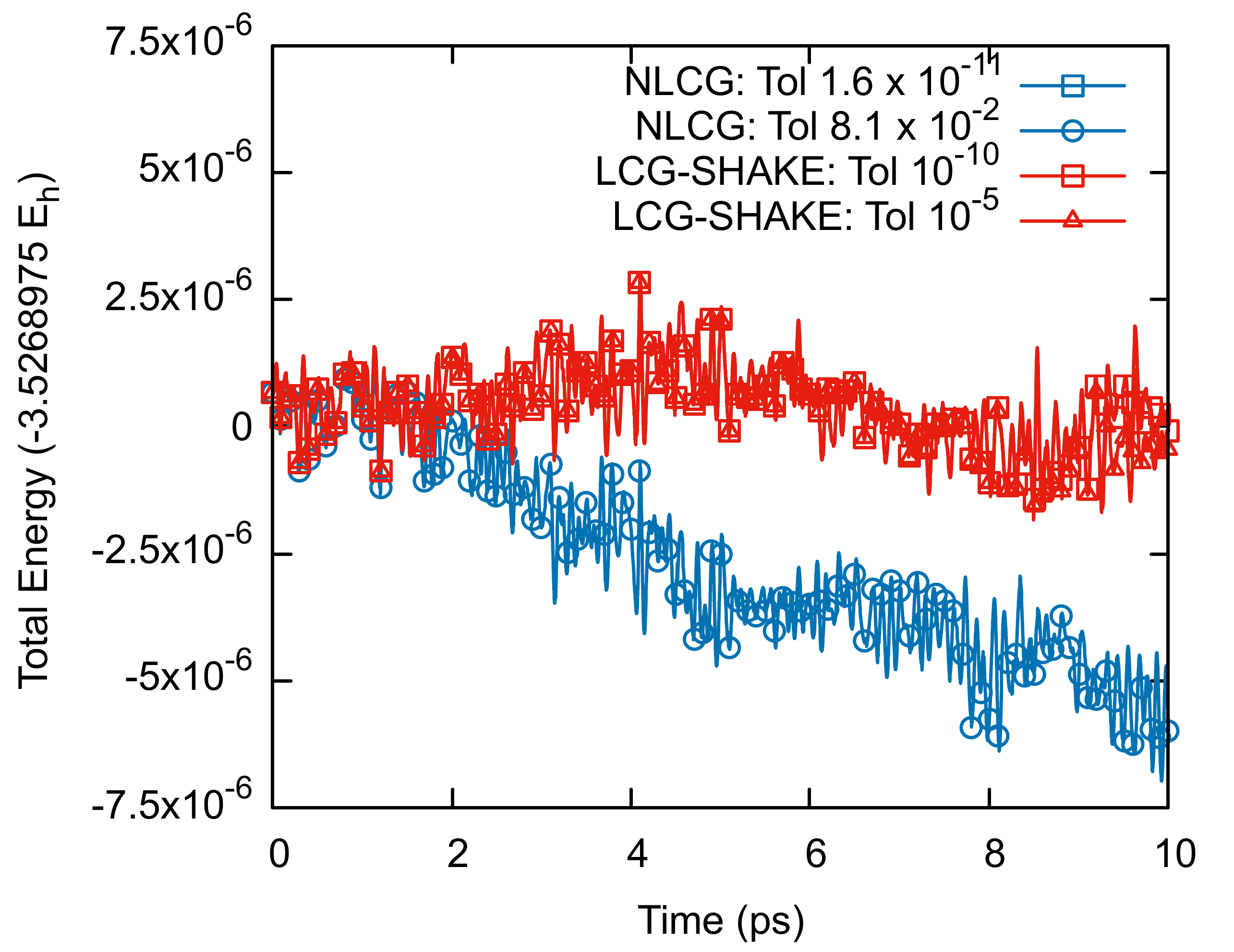}
\end{center}
\caption{Test of energy-conservation properties of MaZe (LCG-SHAKE, curves with red symbols) with respect to standard Born-Oppenheimer alternatives, where the minimization of the energy functional is performed with standard nonlinear conjugate gradient (NLCG, curves with blue symbols) techniques with different convergence thresholds. An intermediate threshold of $10^{-5}$ for MaZe is also presented (red line open triangles) to prove the stability and energy conservation of the algorithm for intermediate tolerances. The total energy on the $y$-axis is shifted by the value at the beginning of the simulation. Results are for the forward trajectories only.} 
\label{fig:EnergyCons}
\end{figure}

Essentially identical results are obtained for energy conservation in the different runs. Figure~\ref{fig:EnergyCons} shows the MaZe (red curve with open squares) and NLCG (blue curve with open squares) energies obtained with the same timestep. Also in this case, when the same stringent convergence threshold is imposed, the results for the two algorithms are superimposed. However, a significant drift in the ionic energy appears as shown (blue curve with open circles) when the convergence in NLCG is relaxed to match the timings of MaZe. 

\section{Conclusions}
This work generalizes previous results for OFDFT \textit{first-principles} dynamics using the MaZe approach in two important ways. Firstly, a detailed derivation of the evolution equations has been presented, highlighting, in particular, the consequences of the complex nature of the Lagrange multipliers associated with the evolution of the coefficients of the expansion of the electronic density in the plane-wave basis. The OFDFT MaZe dynamical system can be integrated combining standard MD algorithms for the evolution of the ions with the SHAKE method to enforce the minimum condition on the electronic degrees of freedom. Efficient implementation of the constraints for generic forms of the kinetic and exchange-correlation functionals, however, is highly nontrivial. We have shown how to achieve linear scaling in the standard implementation of SHAKE for GGA functionals, and introduced an original approach that enables $\Omega \log\Omega$ scaling of the method for arbitrary forms of the kinetic and exchange-correlation functional, including nonlocal expressions where standard SHAKE fails. The accuracy and efficiency of both algorithms has been demonstrated via simulations of a challenging benchmark system, solid and liquid sodium. Use of the MaZe approach beyond constant energy simulations has also been illustrated, presenting results for NVT simulations of the same system.
An in-depth comparison of OFDFT MaZe with alternative state-of-the-art approaches, in particular XL-BOMD and advanced implementations of NLCG Born-Oppenheimer dynamics including preconditioning and smart initialization, will be performed in future studies. Here we make some general preliminary comments. Numerically, when compared to  the original Car-Parrinello dynamics, MaZe has the advantage of enabling the use of a timestep consistent with the ion dynamics instead of the shorter timesteps typically used in Car-Parrinello MD. It also enforces full adiabatic separation via the limit of zero mass for the auxiliary variables. Together with the fact that SHAKE efficiently converges the minimum condition essentially to numerical precision, this prevents energy drifts and avoids the need for thermostats to stabilize the dynamics. Even though the overall approach is different, MaZe shares some similarities the recent XL-BOMD method, such as the presence of a limit of the fictitious mass going to zero. However, in XL-BOMD an extra dissipation of the dynamics of the electronic dofs is needed in practice and the accuracy of propagating the electronic dofs still remains dependent on the integration timestep $\dt$ in XL-BOMD, going as $\dt^{-2}$.\cite{niklasson:2014} MaZe also avoids the numerical cost of computing the kernel involved in XL-BOMD. The advantage over Car-Parrinello and XL-BOMD comes at the non-trivial cost of the computation of the Lagrange multipliers. The calculations presented in this work, however, show that this numerical cost can be controlled and remains quasilinear with respect to the system size.
The developments presented in this paper, and the remarkable efficiency of the algorithms introduced, then open the way to applications of MaZe to more realistic OFDFT models of condensed-phase systems and set the stage to the future extension of the approach to Kohn-Sham based DFT.

\section*{Acknowledgments}
The authors would like to thank Giovanni Ciccotti, Benjamin Rotenberg and Mathieu Salanne for fruitful discussions. Part of this work was supported under the European Union’s Horizon 2020 research and innovation programme under Grant Agreement No. 676531 (project E-CAM). R.V. would also like to acknowledge hospitality at CECAM, where this work started. 

\section*{Data Availability Statement}
The data that support the findings of this study together with the source code used for the numerical simulations are available from the corresponding author upon reasonable request.

\FloatBarrier
\appendix
\begin{widetext}

\section{Expressions of forces  and Hessians and $\Omega\log \Omega$ scaling for matrix-vector multiplications of standard GGA functionals}\label{app:MV_GGA}

The calculations in this paper require the value of the energy functional and its first and second derivatives. The manipulations presented in the main text are for the most part general but in our specific implementation we
exploit the \libxc{} library to obtain these quantities. \libxc{} takes as input the electronic density and its derivatives in real space and returns the value of $\Epsilon$, the energy per unit particle, and its derivatives in real space. The library employs the electronic density $n(\vec{r})$ and the modulus square of its gradient $\varsigma(\vec{r}) = |\nabla n(\vec{r})|^2$ as independent ``variables''. This is the general formulation for a GGA functional. 

The Hartree contribution to first and second derivatives is trivial, the most problematic term being the kinetic-exchange-correlation contribution that we adress first.
The kinetic-exchange-correlation energy is returned as
\begin{equation}
\label{eps_kxc}
E_{\text{KXC}}[n(\vec{r}), \varsigma(\vec{r})] = \frac{\Omega}{\Nd}\sum_{\vec{r}}\varepsilon_{\text{KXC}}\bigl(n(\vec{r}), \varsigma(\vec{r})\bigr),
\end{equation}
where $N_D$ is the number of points in the direct space grid used to represent the discretized density and $\varepsilon_{\text{KXC}}\bigl(n(\vec{r}), \varsigma(\vec{r})\bigr)$ the local kinetic-exchange-energy per unit volume. The library also outputs the derivatives of the functional with respect to the density and to the modulus square of the gradients:
\begin{equation}
\label{eq:v_kxc}
\begin{aligned}
\upsilon_{\text{KXC}}^{n}\bigl(n(\vec{r}), \varsigma(\vec{r})\bigr) &= \frac{\p \varepsilon_{\text{KXC}}\bigl(n(\vec{r}), \varsigma(\vec{r})\bigr)}{\p n(\vec{r})}; \\
\upsilon_{\text{KXC}}^{\varsigma}\bigl(n(\vec{r}), \varsigma(\vec{r})\bigr) &= \frac{\p \varepsilon_{\text{KXC}} \bigl(n(\vec{r}), \varsigma(\vec{r})\bigr)}{\p \varsigma(\vec{r})}.\\
\end{aligned}
\end{equation}
Finally, \libxc{} provides also the following derivatives, associated to the Hessian of the functional
\begin{equation}
\label{eq:f_kxc}
\begin{aligned}
f_{\text{KXC}}^{nn}\bigl(n(\vec{r}), \varsigma(\vec{r})\bigr) &= \frac{\p^2 \varepsilon_{\text{KXC}}\bigl(n(\vec{r}), \varsigma(\vec{r})\bigr)}{\p^2 n(\vec{r})};\\
f_{\text{KXC}}^{n\varsigma}\bigl(n(\vec{r}), \varsigma(\vec{r})\bigr) &= \frac{\p^2 \varepsilon_{\text{KXC}}\bigl(n(\vec{r}), \varsigma(\vec{r})\bigr)}{\p n(\vec{r})\p \varsigma(\vec{r})};\\
f_{\text{KXC}}^{\varsigma\varsigma}\bigl(n(\vec{r}), \varsigma(\vec{r})\bigr) &= \frac{\p^2 \varepsilon_{\text{KXC}}\bigl(n(\vec{r}), \varsigma(\vec{r})\bigr)}{\p^2 \varsigma(\vec{r})}.\\
\end{aligned}
\end{equation}
All derivatives above are defined per unit volume.
The \libxc{} outputs can be easily combined to obtain the first and second derivatives of the kinetic and exchange correlation energy functional with respect to the Fourier coefficients of the electronic densities (see sections~\ref{subsec:StandardSHAKE} and~\ref{subsec:ElberSHAKE}). As an example, we describe in some detail the calculation of the first derivative of the kinetic and exchange functional. We have:
\begin{equation}\label{eq:FirstDer}
\begin{aligned}
\frac{\p E_{\text{KXC}}}{\p \tilde{n}\cc_{\vec{G}_{\alpha}}} &= \frac{\Omega}{N_D}\sum_{\vec{r}}\frac{\p \varepsilon_{\text{KXC}}(\vec{r})}{\p \tilde{n}\cc_{\vec{G}_{\alpha}}} \\ 
& = \frac{\Omega}{N_D}\sum_{\vec{r}}\left[\frac{\p \varepsilon_{\text{KXC}}(\vec{r})}{\p n(\vec{r})}\frac{\p n(\vec{r})}{\p \tilde{n}\cc_{\vec{G}_{\alpha}}} + \frac{\p \varepsilon_{\text{KXC}}(\vec{r})}{\p \varsigma(\vec{r})}\frac{\p \varsigma(\vec{r})}{\p \tilde{n}\cc_{\vec{G}_{\alpha}}}\right] \\
&= \frac{\Omega}{N_D}\sum_{\vec{r}}\left[\frac{\p n(\vec{r})}{\p \tilde{n}\cc_{\vec{G}_{\alpha}}}\upsilon_{\text{KXC}}^{n}(\vec{r}) + \frac{\p \varsigma(\vec{r})}{\p \tilde{n}\cc_{\vec{G}_{\alpha}}}\upsilon_{\text{KXC}}^{\varsigma}(\vec{r})\right],
\end{aligned}
\end{equation}
where the derivatives of the kinetic-exchange-correlation kernel have been defined in Eqs.~\eqref{eq:v_kxc}.
The derivatives of $n(\vec{r})$ and $\varsigma(\vec{r})$ with respect to the coefficients of the electronic density in reciprocal space can be easily computed. From the definition of the Fourier transform of the electronic density, we have
\begin{equation}
\label{eq:dn_dngstar}
\frac{\p n(\vec{r})}{\p \tilde{n}\cc_{\vec{G}_{\alpha}}} = \exp[-\ii\vec{G}_{\alpha}\cdot\vec{r}].
\end{equation} 
The derivative of $\varsigma(\vec{r})$ can be computed using the chain rule 
\begin{equation}
\label{eq:ds_dnstarg}
\frac{\p\varsigma(\vec{r})}{\p \tilde{n}\cc_{\vec{G}_{\alpha}}} = \frac{\p\varsigma(\vec{r})}{\p \grad n(\vec{r})}\cdot\frac{\p\grad n(\vec{r})}{\p \tilde{n}\cc_{\vec{G}_{\alpha}}}.
\end{equation}
Above, the gradient of the density is given by
\begin{equation}
\label{eq:grad_density}
\begin{aligned}
\grad n(\vec{r}) &= \sum^{N_G}_{\beta=1} \Bigl[\tilde{n}_{\vec{G}_{\beta}}\exp[\ii\vec{G}_{\beta}\cdot\vec{r}]\ii\vec{G}_{\beta}-\tilde{n}\cc_{\vec{G}_{\beta}}\exp[-\ii\vec{G}_{\beta}\cdot\vec{r}]\ii\vec{G}_{\beta}\Bigr],
\end{aligned}
\end{equation}
taking the derivative of this quantity with respect to $\tilde{n}\cc_{\vec{G}_{\alpha}}$, and recalling that $\varsigma(\vec{r}) = |\grad n(\vec{r})|^2$, Eq.~\eqref{eq:ds_dnstarg} can then be recast as
\begin{equation}
\label{eq:ds_dngstar}
\frac{\p\varsigma(\vec{r})}{\p \tilde{n}\cc_{\vec{G}_{\alpha}}} = -2\grad n(\vec{r})\cdot\ii\vec{G}_{\alpha}\exp[-\ii\vec{G}_{\alpha}\cdot\vec{r}].
\end{equation}
Substituting Eqs.~\eqref{eq:dn_dngstar} and~\eqref{eq:ds_dngstar} in Eq.~\eqref{eq:FirstDer} we get
\begin{equation}
\begin{aligned}
&\frac{\p E_{\text{KXC}}}{\p \tilde{n}\cc_{\vec{G}_{\alpha}}} = \frac{\Omega}{N_D}\sum_{\vec{r}}\upsilon_{\text{KXC}}^{n}(\vec{r})\exp[-\ii\vec{G}_{\alpha}\cdot\vec{r}] - \frac{\Omega}{N_D}\biggl[2\sum_{\vec{r}}\upsilon_{\text{KXC}}^{\varsigma}(\vec{r})\grad n(\vec{r})\exp[-\ii\vec{G}_{\alpha}\cdot\vec{r}]\biggr]\cdot\ii\vec{G}_{\alpha}.
\end{aligned}
\end{equation}
The expression above can be conveniently rewritten as the sum of two Fourier transforms:
\begin{equation}
\begin{aligned}
\frac{\p E_{\text{KXC}}}{\p \tilde{n}\cc_{\vec{G}_{\alpha}}} &= \frac{\Omega}{N_D}\biggl[\mathrm{FT}\Bigl[\upsilon_{\text{KXC}}^{n}(\vec{r})\Bigr] - 2\mathrm{FT}\Bigl[\upsilon_{\text{KXC}}^{\varsigma}(\vec{r})\grad n(\vec{r})\Bigr]\cdot\ii\vec{G}_{\alpha}\biggr],
\end{aligned}
\end{equation}
and computed numerically via discrete Fourier transforms ($\mathrm{FT}[\bullet]$ in the equation above) with an $\Omega \log \Omega$ cost.

The second derivatives of the kinetic-exchange-correlation functional are  obtained by differentiation of the result above and using, where appropriate, the definitions in Eqs.~\eqref{eq:f_kxc} and the chain rule for the derivatives. For example, we have 
\begin{equation}
\label{eq:constr_gradient_prel}
\begin{aligned}
\frac{\p^2 E_{\text{KXC}}}{\p \tilde{n}_{\vec{G}_{\beta}}\p \tilde{n}\cc_{\vec{G}_{\alpha}}} &= \frac{\Omega}{N_D}\sum_{\vec{r}}\exp[\ii\vec{G}_{\beta}\cdot\vec{r}]\exp[-\ii\vec{G}_{\alpha}\cdot\vec{r}] \\
&\times \biggl[f^{nn}(\vec{r}) - 2\ii \p_l n(\vec{r})(- G_{\beta}^l+ G_{\alpha}^l)f^{n\varsigma}(\vec{r})+2G_{\beta}^lG_{\alpha}^m\bigl(\upsilon^{\varsigma}(\vec{r})\delta_{lm} + 2\p_l n(\vec{r})\p_m n(\vec{r})f^{\varsigma\varsigma}(\vec{r})\bigr)\biggr].
\end{aligned}
\end{equation}
In the equation above, $\p_ln(\vec{r})$ indicates the derivative of the density with respect to the Cartesian component $l$ of the position $\vec{r}$ and repeated Latin indexes are summed over. For future convenience, we define the functions
\begin{equation}
\label{eq:aux_func_hessian}
\begin{aligned}
\Xi(\vec{r}) &\equiv f^{nn}(\vec{r}); \\
\Phi^l(\vec{r}) &\equiv 2 \p_l n(\vec{r})f^{n\varsigma}(\vec{r}); \\
\Psi^{lm}(\vec{r}) &\equiv 2\bigl(\upsilon^{\varsigma}(\vec{r})\delta_{lm} + 2\p_l n(\vec{r})\p_m n(\vec{r})f^{\varsigma\varsigma}(\vec{r})\bigr),
\end{aligned}
\end{equation}
to rewrite Eq.~\eqref{eq:constr_gradient_prel} as
\begin{equation}
\label{eq:sigma_kxc_app}
\begin{aligned}
&\frac{\p^2 E_{\text{KXC}}}{\p \tilde{n}_{\vec{G}_{\beta}}\p \tilde{n}\cc_{\vec{G}_{\alpha}}} = \frac{\Omega}{N_D}\sum_{\vec{r}}\exp[\ii\vec{G}_{\beta}\cdot\vec{r}]\exp[-\ii\vec{G}_{\alpha}\cdot\vec{r}] \times \biggl[ \Xi(\vec{r}) +\ii (G_{\beta}^l- G_{\alpha}^l)\Phi^l(\vec{r})+ G_{\beta}^lG_{\alpha}^m \Psi^{lm}(\vec{r}) 
\biggr].
\end{aligned}
\end{equation}
The second derivatives with respect to different combination of coefficients and complex conjugates of coefficients that are needed in the calculation of the matrix $\Sigma^\dagger= \Sigma^\dagger=\Sigma^\dagger_{\mathrm{Hart}}+\Sigma^\dagger_{\mathrm{KXC}}$ can be obtained from the expression above using the rules of complex conjugation. 

While the Hartree term is diagonal in the density coefficients when using plane waves
\begin{equation}
    \left\{\Sigma^\dagger_{\mathrm{Hart}}\right\}_{\alpha,\beta}=\left\{\Sigma^\dagger_{\mathrm{Hart}}\right\}_{\alpha,N_G+\beta}=\frac{4\pi\Omega}{\vert \vec{G}_\alpha \vert^2}\delta_{\alpha,\beta},
\end{equation}
and so the matrix-vector product is trivial, the final result for the matrix $\Sigma^\dagger_{\mathrm{KXC}}$ is given by
\begin{equation}
\begin{aligned}
    \left\{\Sigma^\dagger_{\mathrm{KXC}}\right\}_{\alpha,\beta}&= \frac{\Omega}{N_D}\sum_{\vec{r}}\exp[\ii\vec{G}_{\beta}\cdot\vec{r}]\exp[-\ii\vec{G}_{\alpha}\cdot\vec{r}] \times \biggl[\Xi(\vec{r}) + \ii (G^l_{\beta} - G^l_{\alpha})\Phi^l(\vec{r})+G^l_{\beta}G^m_{\alpha}\Psi^{lm}(\vec{r})\biggr];\\
    \left\{\Sigma^\dagger_{\mathrm{KXC}}\right\}_{\alpha,N_G+\beta}&= \frac{\Omega}{N_D}\sum_{\vec{r}}\exp[-\ii\vec{G}_{\beta}\cdot\vec{r}]\exp[-\ii\vec{G}_{\alpha}\cdot\vec{r}] \times \biggl[\Xi(\vec{r}) - \ii (G^l_{\beta} + G^l_{\alpha})\Phi^l(\vec{r}))-G^l_{\beta}G^m_{\alpha}\Psi^{lm}(\vec{r})\biggr].
\end{aligned}
\end{equation}
which is more complicated to treat. 
Considering the three terms in the square brackets separately and starting from the first one, the matrix-vector product can however be written
\begin{equation}
\begin{aligned}
&\frac{\Omega}{N_D}\sum_{\beta=1}^{N_G} \sum_{\vec{r}}\exp[\ii\vec{G}_{\beta}\cdot\vec{r}]\exp[-\ii\vec{G}_{\alpha}\cdot\vec{r}] \Xi(\vec{r}) \gamma_{\vec{G}_{\beta}}
   \frac{\Omega}{N_D}\sum_{\beta=1}^{N_G} \sum_{\vec{r}}\exp[-\ii\vec{G}_{\beta}\cdot\vec{r}]\exp[-\ii\vec{G}_{\alpha}\cdot\vec{r}] \Xi(\vec{r}) \gamma\cc_{\vec{G}_{\beta}}
   \\& = 2\Omega\sum_{\vec{r}}\exp[-\ii\vec{G}_{\alpha}\cdot\vec{r}] \Xi(\vec{r}) \mathrm{IFT}[\vec{\gamma}](\vec{r}) = 2\Omega \mathrm{FT}\bigr[\Xi(\vec{r}) \mathrm{IFT}[\vec{\gamma}](\vec{r})\bigl](\vec{G}_\alpha),
\end{aligned}
\end{equation}
where $\mathrm{FT}[\bullet]$ denotes a discrete Fourier transform and $\mathrm{IFT}[\bullet]$ an inverse discrete Fourier transform. Note that the discrete inverse Fourier transform is defined including the term $\vec{G}=\vec{G}_0=(0,0,0)$, which is excluded from the constrained degrees of freedom in our approach. In the numerical implementation of the Fourier transform above, we thus set $\gamma_{\vec{G}_{0}}=0$ to include this term in the sum over the vectors in reciprocal space.

Using Fast Fourier Transforms these operations can thus be done with $\Omega \log \Omega$ scaling. Similar manipulations show that the same is true also for the other two terms.

\section{$\Omega\log \Omega$ scaling for matrix-vector multiplications of nonlocal functionals}\label{app:MV_nonlocal}

For kinetic energy functionals with a nonlocal kernel, we consider the form discussed by Wang et al. for density-independent kernels~\cite{wang:1999}, which we write in the more general form
\begin{equation}
    T_{\mathrm{K}}^{f,g}= C_{\mathrm{K}} \sum_{\vec{r}}\sum_{\vec{r}^\prime}
    f\bigl(n(\vec{r})\bigr)w(\vec{r}-\vec{r}^\prime)g\bigl(n(\vec{r}^\prime)\bigr),
\end{equation}
where $C_K$, is a constant, $f$ and $g$ are generic functions of the density (Wang et al. considered only powers, compare with Eq. (46) of Ref.~\citenum{wang:1999}) and $w(\vec{r}-\vec{r}^\prime)$ is the nonlocal kernel. 
Products of this form are conveniently expressed in Fourier space as
\begin{equation}
    T_{\mathrm{K}}^{f,g}= C_{\mathrm{K}} \sum_{\vec{G}} \widetilde{F}\cc(\vec{G})\widetilde{W}(\vec{G}),
    \widetilde{G}(\vec{G}),
\end{equation}
where $\widetilde{F}(\vec{G})$, $\widetilde{G}(\vec{G})$ and $\widetilde{W}(\vec{G})$ indicates the Fourier transforms of the functions $f\bigl(n(\vec{r})\bigr)$, $g\bigl(n(\vec{r})\bigr)$ and $w(\vec{r})$, respectively. Note that these Fourier transforms include the term $\vec{G}=\vec{G}_0=(0,0,0)$ and the sum extends from $-\vec{G}_\mathrm{max}$ to $\vec{G}_\mathrm{max}$.

Let us begin by considering the first derivative of a kinetic energy functional of this nonlocal form. We have:
\begin{equation}\label{eq:NonLocKinForce}
    \begin{aligned}
    \frac{\p T_{\mathrm{K}}^{f,g}}{\p \tilde{n}\cc_{\vec{G}_\alpha}} &=
    C_{\mathrm{K}} \sum_{\vec{r}}\sum_{\vec{r}^\prime}
    \frac{\p f}{\p n}\bigl(n(\vec{r})\bigr)e^{-\ii \vec{G}_\alpha\vec{r}}w(\vec{r}-\vec{r}^\prime)g\bigl(n(\vec{r}^\prime)\bigr)+C_{\mathrm{K}} \sum_{\vec{r}}\sum_{\vec{r}^\prime}
    f\bigl(n(\vec{r})\bigr)w(\vec{r}-\vec{r}^\prime)\frac{\p g}{\p n}\bigl(n(\vec{r}^\prime)\bigr)e^{-\ii \vec{G}_\alpha\vec{r}^\prime}\\
    &= C_{\mathrm{K}} \sum_{\vec{r}}
    \frac{\p f}{\p n}\bigl(n(\vec{r})\bigr)e^{-\ii\vec{G}_\alpha\vec{r}} \mathrm{IFT}[\widetilde{W}(\vec{G})\widetilde{G}(\vec{G})](\vec{r})+ C_{\mathrm{K}} \sum_{\vec{r}^\prime}
     \mathrm{IFT}[\widetilde{W}\cc(\vec{G})\widetilde{F}(\vec{G})]\cc(\vec{r}^\prime)
     \frac{\p g}{\p n}\bigl(n(\vec{r}^\prime)\bigr)e^{-\ii\vec{G}_\alpha\vec{r}^\prime}\\
     &=C_{\mathrm{K}} \mathrm{FT}\biggl[
      \frac{\p f}{\p n}\bigl(n(\vec{r})\bigr) \mathrm{IFT}[\widetilde{W}(\vec{G})\widetilde{G}(\vec{G})](\vec{r})
    \biggr](\vec{G}_\alpha)+C_{\mathrm{K}} \mathrm{FT}\biggl[
     \mathrm{IFT}[\widetilde{W}\cc(\vec{G})\widetilde{F}(\vec{G})]\cc(\vec{r}^\prime)
     \frac{\p g}{\p n}\bigl(n(\vec{r}^\prime)\bigr)
    \biggr](\vec{G}_\alpha)
    \end{aligned}
\end{equation}
As observed by Wang et al., the evaluation of this term --- which only involves direct and inverse Fourier transforms --- can be performed numerically with $\Omega \log \Omega$ effort. 

We now show that also the product of the matrix of second derivatives of the nonlocal kinetic energy with the $\vec{\gamma}$ needed for the solution of our constraints (see also first part of this Appendix for GGA functionals) can be performed with the same numerical scaling. 
We start from
\begin{equation}
    \begin{aligned}
    \sum_{\beta=1}^{N_G}\biggl\{\Sigma^\dagger_{T}\biggr\}_{\alpha,\beta}\gamma_{\vec{G}_\beta}=\sum_{\beta=1}^{N_G}\frac{\p^2 T_{\mathrm{K}}^{f,g}}{\p \tilde{n}_{\vec{G}_\beta}\p \tilde{n}\cc_{\vec{G}_\alpha}} \gamma_{\vec{G}_\beta} =&
        C_{\mathrm{K}} \sum_{\vec{r}}\sum_{\vec{r}^\prime}
    \frac{\p^2 f}{\p n^2}\bigl(n(\vec{r})\bigr)e^{-\ii \vec{G}_\alpha\vec{r}}\biggl(\sum_{\beta=1}^{N_G}e^{\ii \vec{G}_\beta\vec{r}}\gamma_{\vec{G}_\beta} \biggr)w(\vec{r}-\vec{r}^\prime)g\bigl(n(\vec{r}^\prime)\bigr)\\
    +&C_{\mathrm{K}} \sum_{\vec{r}}\sum_{\vec{r}^\prime}
    \frac{\p f}{\p n}\bigl(n(\vec{r})\bigr)e^{-\ii \vec{G}_\alpha\vec{r}}w(\vec{r}-\vec{r}^\prime)\frac{\p g}{\p n}\bigl(n(\vec{r}^\prime)\bigr)\biggl(\sum_{\beta=1}^{N_G}e^{\ii \vec{G}_\beta\vec{r}^\prime}\gamma_{\vec{G}_\beta} \biggr)\\
    +&C_{\mathrm{K}} \sum_{\vec{r}}\sum_{\vec{r}^\prime}
    \frac{\p f}{\p n}\bigl(n(\vec{r})\bigr)\biggl(\sum_{\beta=1}^{N_G}e^{\ii \vec{G}_\beta\vec{r}}\gamma_{\vec{G}_\beta} \biggr)w(\vec{r}-\vec{r}^\prime)\frac{\p g}{\p n}\bigl(n(\vec{r}^\prime)\bigr)e^{-\ii \vec{G}_\alpha\vec{r}^\prime}\\
    +&    C_{\mathrm{K}} \sum_{\vec{r}}\sum_{\vec{r}^\prime} f\bigl(n(\vec{r})\bigr)
    w(\vec{r}-\vec{r}^\prime)\frac{\p^2 g}{\p n^2}\bigl(n(\vec{r}^\prime)\bigr)\biggl(\sum_{\beta=1}^{N_G}e^{\ii \vec{G}_\beta\vec{r}^\prime}\gamma_{\vec{G}_\beta} \biggr)e^{-\ii \vec{G}_\alpha\vec{r}^\prime}    \end{aligned}
\end{equation}
where we have used the derivative of Eq.~\eqref{eq:NonLocKinForce} with respect to $\tilde{n}_{\vec{G}_\beta}$.
Adding to this the second term of the matrix-vector product $\sum_{\beta=1}^{N_G}\left\{\Sigma^\dagger_{T}\right\}_{\alpha,N_G+\beta}\gamma\cc_{\vec{G}_\beta}$
and defining $\gamma_D(\vec{r})=\sum_{\beta=1}^{N_G}e^{\ii \vec{G}_\beta\vec{r}}\gamma_{\vec{G}_\beta}+\sum_{\beta=1}^{N_G}e^{-\ii \vec{G}_\beta\vec{r}}\gamma\cc_{\vec{G}_\beta}=\mathrm{FT}[\gamma](\vec{r})$ (where again, in the numerical implementation of the Fourier transform FT, we set $\gamma_{\vec{G}_0}=0$), we can rewrite the matrix-vector product  as
\begin{equation}
    \begin{aligned}
\sum_{\beta=1}^{N_G}\left\{\Sigma^\dagger_{T}\right\}_{\alpha,\beta}\gamma_{\vec{G}_\beta}
+
\sum_{\beta=1}^{N_G}\left\{\Sigma^\dagger_{T}\right\}_{\alpha,N_G+\beta}\gamma\cc_{\vec{G}_\beta}
&=
        C_{\mathrm{K}} \sum_{\vec{r}}\sum_{\vec{r}^\prime}
    \frac{\p^2 f}{\p n^2}\bigl(n(\vec{r})\bigr)e^{-\ii \vec{G}_\alpha\vec{r}}\gamma_D(\vec{r})w(\vec{r}-\vec{r}^\prime)g\bigl(n(\vec{r}^\prime)\bigr)\\
    +&C_{\mathrm{K}} \sum_{\vec{r}}\sum_{\vec{r}^\prime}
    \frac{\p f}{\p n}\bigl(n(\vec{r})\bigr)e^{-\ii \vec{G}_\alpha\vec{r}}w(\vec{r}-\vec{r}^\prime)\frac{\p g}{\p n}\bigl(n(\vec{r}^\prime)\bigr)\gamma_D(\vec{r}^\prime)\\
    +&C_{\mathrm{K}} \sum_{\vec{r}}\sum_{\vec{r}^\prime}
    \frac{\p f}{\p n}\bigl(n(\vec{r})\bigr)\gamma_D(\vec{r})w(\vec{r}-\vec{r}^\prime)\frac{\p g}{\p n}\bigl(n(\vec{r}^\prime)\bigr)e^{-\ii \vec{G}_\alpha\vec{r}^\prime}\\
    +&    C_{\mathrm{K}} \sum_{\vec{r}}\sum_{\vec{r}^\prime} f\bigl(n(\vec{r})\bigr)
    w(\vec{r}-\vec{r}^\prime)\frac{\p^2 g}{\p n^2}\bigl(n(\vec{r}^\prime)\bigr)\gamma_D(\vec{r}^\prime)e^{-\ii \vec{G}_\alpha\vec{r}^\prime}.    
    \end{aligned}
\end{equation}
The structure of the four terms in the expression above is the same as that of the two terms involved in the expression of the first derivative of the kinetic energy functional, Eq.~\eqref{eq:NonLocKinForce}. Using the same manipulations, each one of them can then be expressed as a sequence of direct and inverse Fourier transforms, thus leading to an $\Omega \log \Omega$ numerical effort in the calculations. 

\section{$\Omega\log\Omega$ scaling for computation of the SHAKE matrix for GGA functionals}
\label{app:SM_GGA}
In the following we show how the cost of computing the diagonal elements of the SHAKE matrix (see eq.~\eqref{eq:SHAKEDenPrestigeStart}) can be reduced to $\Omega \log\Omega$ scaling by exploiting periodicity in direct space. We report below the expression of these diagonal elements for convenience: 
\begin{equation*}
\begin{aligned}
\left[\mathbb{J}_D\right ]_{{\vec{G}_{\alpha}},\vec{G}_{\alpha}}   =&\left(\frac{4\pi\Omega}{|\vec{G}_{\alpha}|^2}\right)^2+ \frac{4\pi\Omega}{|\vec{G}_{\alpha}|^2} \frac{\Omega}{N_D} \sum_{\vec{r}}\Bigl[\Xi_{\text{P}}(\vec{r}) + G^l_{\alpha}G^m_{\alpha}\Psi_{\text{P}}^{lm}(\vec{r})\Bigr]+ \frac{4\pi\Omega}{|\vec{G}_{\alpha}|^2} \frac{\Omega}{N_D} \sum_{\vec{r}}\Bigl[\Xi_{t}(\vec{r}) + G^l_{\alpha}G^m_{\alpha}\Psi_{t}^{lm}(\vec{r})\Bigr]\\
&+ \biggl(\frac{\Omega}{N_D}\biggr)^2\sum_{\vec{r},\vec{r}'}e^{-\ii\vec{G}_{\alpha}\cdot\vec{r}}e^{\ii\vec{G}_{\alpha}\cdot\vec{r}'}\sum^{N_G}_{\lambda=1}e^{\ii\vec{G}_{\lambda}\cdot\vec{r}}e^{-\ii\vec{G}_{\lambda}\cdot\vec{r}'} \\
&\qquad\times \Bigl[\Xi_{\text{P}}(\vec{r}) + \ii (G^l_{\lambda} - G^l_{\alpha})\Phi^{l}_{\text{P}}(\vec{r}) + G^l_{\lambda}G^m_{\alpha}\Psi_{\text{P}}^{lm}(\vec{r})\Bigr]
\Bigl[\Xi_{t}(\vec{r}') - \ii (G^l_{\lambda} - G^l_{\alpha})\Phi_{t}^l(\vec{r}' ) + G^l_{\lambda}G^m_{\alpha}\Psi_{t}^{lm}(\vec{r}')\Bigr] \\
&+ 
\biggl(\frac{\Omega}{N_D}\biggr)^2\sum_{\vec{r},\vec{r}'}e^{-\ii\vec{G}_{\alpha}\cdot\vec{r}}e^{\ii\vec{G}_{\alpha}\cdot\vec{r}'}\sum^{N_G}_{\lambda=1}e^{-\ii\vec{G}_{\lambda}\cdot\vec{r}}e^{\ii\vec{G}_{\lambda}\cdot\vec{r}'} \\
&\qquad\times \Bigl[\Xi_{\text{P}}(\vec{r}) - \ii (G^l_{\lambda} + G^l_{\alpha})\Phi_{\text{P}}^l(\vec{r}) - G^l_{\lambda}G^m_{\alpha}\Psi_{\text{P}}^{lm}(\vec{r})\Bigr]
\times \Bigl[\Xi_{t}(\vec{r}') + \ii (G^l_{\lambda} + G^l_{\alpha})\Phi_{t}^l(\vec{r}' ) - G^l_{\lambda}G^m_{\alpha}\Psi_{t}^{lm}(\vec{r}')\Bigr],
\end{aligned}
\end{equation*}
which we rewrite as
\begin{equation}
\label{eq:JKXC}
\begin{aligned}
\left[\mathbb{J}_D\right ]_{{\vec{G}_{\alpha}},\vec{G}_{\alpha}} = \left(\frac{4\pi\Omega}{|\vec{G}_{\alpha}|^2}\right)^2 + \frac{4\pi\Omega}{|\vec{G}_{\alpha}|^2} \frac{\Omega}{N_D} \biggl(\sum_{\vec{r}}\Bigl[\Xi_{\text{P}}(\vec{r}) + G^l_{\alpha}G^m_{\alpha}\Psi_{\text{P}}^{lm}(\vec{r})\Bigr] + \sum_{\vec{r}}\Bigl[\Xi_{t}(\vec{r}) + G^l_{\alpha}G^m_{\alpha}\Psi_{t}^{lm}(\vec{r})\Bigr]\biggr) + \left[\mathbb{J}^{\text{KXC}}_D\right ]_{{\vec{G}_{\alpha}},\vec{G}_{\alpha}} 
\end{aligned}
\end{equation}
The first term involving the Hartree contribution, as well as the mixed terms, have a numerical cost that trivially scales linearly with system size.
We will discuss in this appendix the pure kinetic and exchange-correlation contribution $\left[\mathbb{J}^{\text{KXC}}_D\right ]_{{\vec{G}_{\alpha}},\vec{G}_{\alpha}}$.
Let us begin by observing that performing the products in the triple sum above results in a total of eight terms, all of the generic form
\begin{equation}
\begin{aligned}
h^{ab}_1(\vec{G}_{\alpha}) &= \sum_{\vec{r}' = 0}^{\vec{L}}\sum_{\vec{r} = 0}^{\vec{L}}\sum_{\lambda=1}^{N_G}\biggl[ e^{-\ii\vec{G}_{\alpha}\cdot\vec{r}}e^{\ii\vec{G}_{\lambda}\cdot\vec{r}}e^{\ii\vec{G}_{\alpha}\cdot\vec{r}'}e^{-\ii\vec{G}_{\lambda}\cdot\vec{r}'} \times (\ii G^l_{\lambda})^a (\ii G^m_{\lambda})^b \phi(\vec{r})\psi(\vec{r}')\biggr]
\end{aligned}
\end{equation}
where we have explicitly indicated the range of the sums in direct space, $a,b=0,1$ and the functions $\phi$ and $\psi$ are proxies for the $\Xi$ and $\Phi$ computed at iteration $\kappa$ or at time $t$ values of the electronic coefficients. Focusing at first on the sum over $\vec{r}$, note that, since both $\Phi$ and $\Xi$ are functions of the (periodic) energy of the system, the function $\phi(\vec{r})e^{-\ii\vec{G}_{\alpha}\cdot\vec{r}}e^{\ii\vec{G}_{\lambda}\cdot\vec{r}}$ is periodic in $\vec{r}$ with period $\vec{L}$. The sums in the equation above can then be reorganized as follows 
\begin{equation}
\begin{aligned}
h^{ab}_1(\vec{G}_{\alpha}) &= \sum_{\vec{r}' = 0}^{\vec{L}}\sum_{\vec{r} = \vec{r}'}^{\vec{L}+\vec{r}'}\sum_{\lambda=1}^{N_G}\biggl[ e^{-\ii\vec{G}_{\alpha}\cdot\vec{r}}e^{\ii\vec{G}_{\lambda}\cdot\vec{r}}e^{\ii\vec{G}_{\alpha}\cdot\vec{r}'}e^{-\ii\vec{G}_{\lambda}\cdot\vec{r}'} \times (\ii G^l_{\lambda})^a (\ii G^m_{\lambda})^b \phi(\vec{r})\psi(\vec{r}')\biggr]
\end{aligned}
\end{equation}
from which, using the transformation $\vec{r} = \vec{r}'+\vec{r}''$,
\begin{equation}
\label{eq:h1_after_sub}
\begin{aligned}
h^{ab}_1(\vec{G}_{\alpha}) &= \sum_{\vec{r}' = 0}^{\vec{L}}\sum_{\vec{r}'' = 0}^{\vec{L}}\sum_{\lambda=1}^{N_G}\biggl[ e^{-\ii\vec{G}_{\alpha}\cdot\vec{r}''}e^{\ii\vec{G}_{\lambda}\cdot\vec{r}''}\times (\ii G^l_{\lambda})^a (\ii G^m_{\lambda})^b \phi(\vec{r}'+\vec{r}'')\psi(\vec{r}')\biggr]
\end{aligned}
\end{equation}
The terms in Eq.~\eqref{eq:h1_after_sub} can now be rearranged to obtain
\begin{equation}
\label{eq:h1_appear_corr}
\begin{aligned}
h^{ab}_1(\vec{G}_{\alpha}) &= \sum_{\vec{r}'' = 0}^{\vec{L}} e^{-\ii\vec{G}_{\alpha}\cdot\vec{r}''}\biggl[\biggl(\sum_{\lambda=1}^{N_G}e^{\ii\vec{G}_{\lambda}\cdot\vec{r}''}
(\ii G^l_{\lambda})^a (\ii G^m_{\lambda})^b \biggr)\biggl(\sum_{\vec{r}' = 0}^{\vec{L}}\phi(\vec{r}'+\vec{r}'')\psi(\vec{r}')\biggr)\biggr]
\end{aligned}
\end{equation}
The sum over $\vec{r}'$ in the equation above is, by definition, the spatial correlation function $\left(\phi\star\psi\right)(\vec{r}'')=\sum_{\vec{r}'= 0}^{\vec{L}}\phi(\vec{r}'+\vec{r}'')\psi(\vec{r}')$. 

The second triple sum in Eq.~\eqref{eq:SHAKEDenPrestigeStart} can be treated in the same way to obtain
\begin{equation}
\begin{aligned}
h^{ab}_2(\vec{G}_{\alpha}) &= \sum_{\vec{r}'' = 0}^{\vec{L}} e^{-\ii\vec{G}_{\alpha}\cdot\vec{r}''}\biggl[\biggl(\sum_{\lambda=1}^{N_G}e^{-\ii\vec{G}_{\lambda}\cdot\vec{r}''}
(-\ii G^l_{\lambda})^a (-\ii G^m_{\lambda})^b \biggr)\biggl(\sum_{\vec{r}' = 0}^{\vec{L}}\phi(\vec{r}'+\vec{r}'')\psi(\vec{r}')\biggr)\biggr]
\end{aligned}
\end{equation}
Combining these results, we have
\begin{equation}
h^{ab}_1(\vec{G}_{\alpha}) + h^{ab}_2(\vec{G}_{\alpha})
= 
\sum_{\vec{r}'' = 0}^{\vec{L}}\left(\phi\star\psi\right)(\vec{r}'')e^{-\ii\vec{G}_{\alpha}\cdot\vec{r}''}
\sum_{\lambda=1}^{N_G}\biggl[
e^{\ii\vec{G}_{\lambda}\cdot\vec{r}''}
(\ii G^l_{\lambda})^a (\ii G^m_{\lambda})^b 
+
e^{-\ii\vec{G}_{\lambda}\cdot\vec{r}''}
(-\ii G^l_{\lambda})^a (-\ii G^m_{\lambda})^b 
\biggr]
\end{equation}
The expression above is particularly convenient for our purposes because the remaining sums over direct and reciprocal space are all, directly or via trivial manipulations, direct or inverse Fourier transforms that can be computed at $\Omega \log \Omega$ (or equivalently $N_D\log N_D$) cost using libraries implementing the fast Fourier transform (FFT) algorithm. More in detail, the sum over $\vec{G}_{\lambda}$ is, by definition, the inverse (discrete) Fourier transform of the product $(iG^l_{\lambda})^a(iG^m_{\lambda})^b$ multiplied by $N_D$, while the correlation function $\left(\phi\star\psi\right)$ can be computed as
\begin{equation}
\left(\phi\star\psi\right)(\vec{r}'') = \sum_{\vec{r}'}\phi(\vec{r}' + \vec{r}'')\psi(\vec{r}') = \Nd\mathrm{IFT}\bigl[\mathrm{FT}[\phi]\cc\mathrm{FT}[\psi]\bigr]
\end{equation}
via the convolution theorem.  Thus,
\begin{equation}\label{eq:theprestige}
\begin{aligned}
h^{ab}_1(\vec{G}_{\alpha}) + h^{ab}_2(\vec{G}_{\alpha}) & =  N^2_D\sum_{a,b}\FFT\biggl[\biggl(\iFFT[(\ii G^l_{\lambda})^a (\ii G^m_{\lambda})^b]\biggr)\biggl(\iFFT\bigl[\FFT[\phi]\cc\FFT[\psi]\bigr]\biggr)\biggr]
\end{aligned}
\end{equation}
Note that the FTs in Eq.~\eqref{eq:theprestige} return a vector of length $N_D$ whose components are the $N_D$ nonzero elements of $h^{ab}_1 + h^{ab}_2$. This implies that, together with the other terms in Eq.~\eqref{eq:JKXC}, the calculation of the SHAKE denominator is reduced to a number of $\mathcal{O}(\Nd\log\Nd)$ operations equal to the number of terms of the form of the LHS of Eq.~\eqref{eq:theprestige}.
Finally, the diagonal part of the KXC contribution to the hessian, $\mathbb{J}^{\text{KXC}}_D$, is obtained as a linear combination of these terms $h^{ab}_1 + h^{ab}_2$, with some coefficients involving a product with $G^l_{\alpha}$ or $G^m_{\alpha}$. This however comes at nearly no extra cost and is in any case linear in the system size so that the overall cost proportional to $\mathcal{O}(\Nd\log\Nd)$ is achieved.
\end{widetext}

\end{document}